\documentclass{elsarticle}
\usepackage{amsfonts,amssymb,amsmath,latexsym,epsfig}
\usepackage{graphicx}
\newcommand{\gsim}{\raisebox{0.8mm}{\em $\,>$}\hspace{-2.45mm}\raisebox{-0.9mm}{\em $\sim \,$}}
\begin{document}

\begin{frontmatter}
\title{Switching between oscillations and homeostasis in competing negative
and positive feedback motifs}

\author[label4,label1]{Weihan Li} \address[label4]{Niels Bohr
  Institute, University of Copenhagen, Blegdamsvej 17, 2100-DK,
  Copenhagen, Denmak.  }
\address[label1]{ Department of Biochemistry and
  Biophysics, University of California San Francisco, San Francisco,
  CA 94158, USA.  } 
\author[label2]{Sandeep Krishna} \address[label2]{National Centre
  for Biological Sciences, Bangalore, Karnataka, India.  }
\author[label3]{Simone Pigolotti} \address[label3]{Dept. de Fisica i Eng. Nuclear,
  Universitat Politecnica de Catalunya Edif. GAIA, Rambla Sant Nebridi
  s/n, E-08222 Terrassa, Barcelona, Spain.  } 
\author[label4]{Namiko Mitarai}
\author[label4]{Mogens H. Jensen
\corref{cor1}
}\ead{mhjensen@nbi.dk} 
\cortext[cor1]{Corresponding author.} % Tel/Fax:+45 35325425.} 

\begin{abstract}
  We analyze a class of network motifs in which a short, two-node
  positive feedback motif is inserted in a three-node negative
  feedback loop.  We demonstrate that such networks can undergo a
  bifurcation to a state where a stable fixed point and a stable limit
  cycle coexist.  At the bifurcation point the period of the
  oscillations diverges.  Further, intrinsic noise can make the system
  switch between oscillatory state and the stationary state
  spontaneously. We find that this switching also occurs in previous
  models of circadian clocks that use this combination of positive and
  negative feedback. Our results suggest that real-life circadian
  systems may need specific regulation to prevent or minimize such
  switching events.
\end{abstract}
\end{frontmatter}

\section{Introduction}

Genetic regulatory networks exhibit a wide range of oscillatory
phenomena, ranging from the fast oscillations in calcium
\cite{Schuster} to the 24 hour cycle of circadian clocks
\cite{Winfree,Marc}.  The occurrence of oscillations is generally
caused by the presence of a negative feedback loop in the regulatory
network \cite{Tiana,Pigolotti}. From a theoretical point of view,
negative feedback can cause a Hopf bifurcation and thus generate a
transition between a stable fixed point, corresponding to homeostasis,
and an attracting limit cycle, corresponding to oscillations.

However, in real regulatory networks, the loop causing oscillations is
usually embedded in a larger network including multiple positive and
negative feedbacks. In some cases, these additional loops have a
demonstrable biological function, for example in 
giving tunability to the oscillation period \cite{Tsai} or in
stabilizing the
period of circadian clocks in the presence of temperature fluctuations
or molecular noise \cite{Gonze}. In general, one expects that multiple
feedback loops could lead to non-trivial behaviors from the viewpoint
of bifurcation theory. The dynamics could become even richer when the
noise induced by stochastic gene expression is taken into account. As
we discuss later, circadian clocks are typical examples of genetic
circuits where it may be important to understanding these effects.

In this paper, we present and analyze a class of network motifs in
which a two-node positive feedback motif is inserted into a three-node
negative feedback loop, as represented in Fig. \ref{pattern}.  In
section 2.1, we show that a deterministic dynamical models of the
simplest of such networks, in a suitable parameter range, exhibits
co-existence of a stable fixed point and a stable limit cycle. We
explain this behavior in terms of a saddle-node separatrix-loop
bifurcation \cite{Magnitskii}, and show that it results in a
diverging oscillation period close to the bifurcation point. In
section 2.2, we use stochastic simulations using the Gillespie
algorithm \cite{Gillespie} to demonstrate that the noise can make the
system switch between oscillatory state and the stationary state. We
show that similar behaviour occurs in more realistic models of
circadian clocks that contain the same combination of positive and
negative feedback loops. Section 3 summarizes our thoughts on the
relevance of these results for the behaviour of circadian clocks.

\section{Results}
\subsection{Dynamics of the Network Motif}\label{secdet}

We study the class of genetic networks represented in
Fig. ~\ref{pattern}.  In each of the four networks, a positive
feedback between node 1 and node 2 can give rise to a bi-stable
switch.  When node 3 is introduced, node 1, 2, 3 together form a
negative feedback loop. The negative feedback loop tends to
destabilize one of the stable fixed points of the switch.  The
simplest motifs that exhibit such ``frustrated bistability'' are
studied in ref. \cite{frustbistability}. Here, we study slightly
larger networks which allow for more intricate dynamics, in particular
a scenario where a stable limit cycle emerges around the unstable
fixed point while the other stable fixed point remains unchanged.

We shall first focus our discussion on the network (a) in Fig
~\ref{pattern}, for which we write the following dynamical equations:

\begin{eqnarray}
\frac{dx_1}{dt}&=&c+ \frac{\alpha }{1+({\frac{x_3}{k}})^h} 
\frac{1}{1+({\frac{x_2}{\beta k}})^h }-\gamma x_1,
\label{eq1}\\
\frac{dx_2}{dt}&=&c+ \frac{\alpha }{1+({\frac{x_1}{k}})^h}-\gamma x_2,
\label{eq2}\\
\frac{dx_3}{dt}&=&c+ \frac{\alpha }{1+({\frac{x_2}{k}})^h}-\gamma x_3,
\label{eq3}
\end{eqnarray}

Here, $x_{1,2,3}$ are the concentrations of the proteins associated
with the three nodes, $\alpha$ is the strength of the three inhibitory
regulations, $h$ is the Hill coefficient, $c$ is a constant source
term for each node, and $\gamma$ is the degradation rate for each
protein (we assume that all three proteins are stable, and therefore
their degradation rate is determined by the cell division time).  
$\beta$ is the control parameter to adjust the inhibition from 
node 2 to node 1.
To
simplify the equations, we introduce dimensionless parameters and
variables $X_{1,2,3}= \frac{ x_{1,2,3} }{k} $, $\tau= \gamma t$, $A=
\frac{\alpha }{k \gamma} $, $B= \frac{c }{k \gamma} $:

\begin{eqnarray}
\frac{dX_1}{d\tau}&=&B+ A\frac{1}{1+{X_3}^h} 
\frac{1}{1+({\frac{X_2}{\beta}})^h }- X_1\label{eq4}\\
\frac{dX_2}{d\tau}&=&B+A \frac{1}{1+{X_1}^h}-X_2\label{eq5}\\
\frac{dX_3}{d\tau}&=&B+A \frac{1}{1+{X_2}^h}-X_3 \label{eq6}.
\end{eqnarray}

We study this system of equations for parameter values of $h=3, B=0.1,
A=5$, and vary $\beta$ as a control parameter, which changes the
strength of the positive feedback relative to the negative feedback.
We found three bifurcation points, at $\beta=\lambda_3\approx 2.03$,
$\beta=\lambda_2\approx 1.923$, and $\beta=\lambda_1\approx 1.45$.
When $\beta>\lambda_3$ (Fig.\ref{phase_space}A), only one unstable
fixed point and one stable limit cycle are found in the phase
space. Here, the stable limit cycle is a global attractor.  At
$\beta=\lambda_3$, a
saddle-node bifurcation occurs, where a stable
fixed point and a saddle node emerge. 

When $\lambda_2<\beta<\lambda_3$ (Fig.\ref{phase_space}B), we find one
stable limit cycle and three fixed points -- one stable and two
unstable. Within this region, the stable limit cycle is not the global
attractor. Depending on the initial condition, the system may either
reach the limit cycle or the stable fixed point. In the parameter
range we studied, the volumes of the two basins of attraction are both 
non-negligible, 
separated by the surface shown in Fig ~\ref{basin}.

Approaching the critical point $\lambda_2$, the stable limit cycle
approaches the saddle point, and eventually at $\beta=\lambda_2$
(Fig.\ref{phase_space}C), they generate a homoclinic cycle. When
$\beta$ is near $\lambda_2$, the period of oscillation increases
dramatically, and eventually diverges at the critical point
$\beta=\lambda_2$ (Fig ~\ref{period}) as typical for homoclinic
cycles.  Such a bifurcation is classified as saddle-node
separatrix-loop bifurcations \cite{Magnitskii, Izhikevich} and is a
robust bifurcation scenario in a phase space of dimension $3$ or more
\cite{Magnitskii}. It corresponds to an attractor crisis, so that when
$\beta$ is reduced further, the stable limit cycle disappears.  When
$\lambda_1<\beta<\lambda_2$ (Fig.\ref{phase_space}D), three fixed
points still exist in the phase space: a stable fixed point, a saddle
node and an unstable fixed point with complex eigenvalues.  The stable
fixed point is the global attractor. At $\beta=\lambda_1$, the saddle
node and the unstable fixed point collide and disappear after a
saddle-node bifurcation. So, when $\beta<\lambda_1$
(Fig.\ref{phase_space}E), there is only one fixed point, which is
stable and is a global attractor.

The location and nature of the fixed points can be better understood
by a graphical study of the intersections of the nullclines. We set
$d/d\tau$ terms in equations (\ref{eq4})-(\ref{eq6}) to zero, and
rearrange the resulting algebraic relations to express $X_2$ and $X_3$
in terms of $X_1$. Then, using this to eliminate $X_2$ and $X_3$ in
equation (\ref{eq4}) yields:

\begin{equation}
0=\frac{A}{1+({\frac{A}{1+({\frac{A}{1+X_1^h}+B})^h}+B})^h}
\frac{1}{1+\frac{A}{\beta({1+X_1^h})+\beta B}}+B-X_1.
\label{Eq:one_d}
\end{equation}

Fig ~\ref{phase_space} right column shows the right hand side of
Eq. (\ref{Eq:one_d}) as the parameter $\beta$ is varied.  When
$\beta>\lambda_3$ (Fig.\ref{phase_space}A) , the function has one
zero, corresponding to a unique unstable fixed point in phase
space. When $\lambda_2<\beta<\lambda_3$ (Fig.\ref{phase_space}B),
$\beta = \lambda_2$ (Fig.\ref{phase_space}C) and
$\lambda_1<\beta<\lambda_2$ (Fig.\ref{phase_space}D), the function has
two additional zeroes, indicating two more fixed points. From the
eigen values of all three fixed points, one can show that only one of
them (the circular one) is stable. Finally, when $\beta<\lambda_1$
(Fig.\ref{phase_space}E), the function has one zero again. Two fixed
points disappear and only one stable fixed point remains.

We checked numerically the parameter range where
the same bifurcations and
qualitatively the same phase space portrait 
can be obtained. For $A=5$ and $B=0.1$, 
$h\gsim 3$ is required to see the same behavior.
For $h=3$ and $B=0.1$, $A\gsim 3.5$ is required.
 The behavior is 
found to be insensitive to the value of $B$;
for $h=3$ and $A=5$, the behavior was unchanged
for $0\le B\le 100$.
The same bifurcations can also be found in all the other 3
motifs listed in Fig ~\ref{pattern}. 
Namely, the observed sequence of bifurcations is a robust feature of 
such motifs.

\subsection{Stochastic Dynamics}\label{stochastic}
In this section, we investigate the effect of the intrinsic noise 
due to the discrete nature of molecular reactions 
in such motifs. We use
the Gillespie algorithm \cite{Gillespie} for stochastic simulations
of the dynamical system specified by equations (\ref{eq1})--(\ref{eq3}).
We denote the copy number of 
molecules of species $i$ as $N_i$. The allowed transitions, along with
their kinetic rates, are:
\begin{eqnarray}
&& N_1 \xrightarrow{cV+\alpha V/[(1+(N_3/k)^h)(1+(N_2/\beta k)^h)]} N_1+1,
\label{stoc1}\\
&& 
N_1 \xrightarrow{\gamma N_1} N_1-1,\\
&&  N_2 \xrightarrow{cV+\alpha V/[(1+(N_1/k)^h)]} N_2+1,\\
&&N_2 \xrightarrow{\gamma N_2} N_2-1,\\
&&  N_3 \xrightarrow{cV+\alpha V/[(1+(N_2/k)^h)]} N_3+1,\\
&&N_3 \xrightarrow{\gamma N_3} N_3-1.
\label{stocN}
\end{eqnarray}

To control the noise, we change the volume of the system $V$, which
changes the production rates of $N_i$, but leaves the average
concentration $x_i\equiv N_i/V$ constant, as long as the values of $k,
\gamma, h, c, \alpha$, and $\beta$ are unchanged.  The larger the
value of $V$, and therefore the larger the copy numbers $N_i$, the
smaller the noise (the relative fluctuations in $N_i$).  Note that we
do not explicitly consider processes like mRNA production, binding of
transcription factors, etc. Inclusion of these steps can increase the
noise in the system further \cite{Loinger}.
 
Figure \ref{noise_3_node} shows the concentration $N_i/V$ vs. time 
for a stochastic simulation in dimensionless units with
%$k=1, \gamma=1, h=3, c=0.1, \alpha=1, \beta=2$ 
$h=3, B=0.1, A=5, \beta=2$.  We convert numbers so that the
dimensionless concentration $X_i=1$ corresponds to one molecule when
volume $V=1$, and simulated (A) $V=1000$ and (B)
$V=100$. 
We can clearly see switching between the oscillatory state (the stable
limit cycle) and and the steady state (the stable fixed point),
and this switching happens more often for smaller $V$, 
i.e., for larger noise.

To quantify this switching behavior, we measured the average switching
time from the oscillatory state to the steady state and vice versa as
a function of the system volume $V$ (Fig.\ref{switchrate}).  
Because of the noisy dynamics, switching is  determined by using
 two thresholds for the distance 
from the stable fixed point, $S_1$ and $S_2>S_1$: 
We define a
switching event from the steady regime to osillatory regime when
when the distance exceeds $S_2$,
while the reverse switching happens when 
the distance becomes smaller than $S_1$.
Therefore, when the distance is between $S_1$ and $S_2$,
there is a history dependence in which regime the state
belongs to.
For this
parameter set, the oscillation period is $\sim$8.15, thus we can see
that the system with hundreds of molecules (corresponds to $V\approx
100 $) can still cause frequent switching, on the order of once in
every 10 oscillations.  The switching rate decreases with $V$ as
expected.  For large enough $V$, the switching rate decreases
exponentially with $V$.

Finally, in order to see whether such switching behavior 
can be relevant for real biological systems,
we study effect of noise on the more realistic model for Drosophila
circadian rhythms described in ref. \cite{Leloup1,Leloup2}.  
The deterministic version of this model has a combination
of several positive and negative feedback loops and exhibits 
the coexistence of a stable fixed point and 
a stable limit cycle \cite{Leloup1}.
We simulated the stochastic version of this model
with parameters used in ref. \cite{Leloup2}.
A detailed description of the model and parameters are given in the appendix.
We observe again the switching between the oscillatory state and 
the steady state due to the noise
(Fig ~\ref{noise_Drosophila}). 
The switching is quite often compared to the 
oscillation frequency for the 
noise level expected for an 
average cell volume, i.e. $10^3 \mu m$ (Fig.~\ref{noise_Drosophila}C),
where 1 [nM] corresponds to about 600 molecules\footnote{
Calculated based on 1 nM $\sim 6 \times 10^{23} \times 10^{-9}$ molecules 
per liter.}. 

\section{Summary and Discussion}\label{summary}
We have analyzed a class of network motifs consisting of a two-node
positive feedback inserted into a three-node negative feedback loop.
We demonstrated that a stable fixed point and a stable limit cycle can
co-exist in this class of motifs. As parameters are changed, the
system undergoes a saddle-node separatrix-loop bifurcation, with a
diverging oscillation period as the system approaches the
bifurcation. The location and nature of the fixed points were
investigated in detail by outlining the intersections of the
nullclines of the three variables.  We then studied the effect of
intrinsic noise to the motif, in the parameter regime where a fixed
point and stable limit cycle co-exist.  We showed that stochastic
switching between the two happens with a rate that decreases with 
decreasing the level of the noise.
Actually, our results complement the study in \cite{Kuni}, 
where similar motifs for 
biochemical systems are studied 
focusing on the how a negative feedback perturbs the switch by positive feedback loops.

We further showed that similar behaviour is observed in a more complex
model of the Drosophila circadian clock.  This switching behaviour was
not reported in a study of the effect of noise in a simplified version
of the detailed model \cite{Gonze}.  The studies on the Drosophila
model \cite{Leloup1,Leloup2,Gonze} focused instead on the singular
behaviour of the real circadian clock, where a single pulse of light
can cause long-term suppression of the circadian rhythm
\cite{Winfree,Honma,Leloup1}, The models explain this as a switch from
the limit cycle to the stable fixed point, caused by a short external
perturbation (the pulse of light) to some parameter values. Recent
research suggests an alternate possibility, where the singular
behaviour is caused by desynchronization of the clocks rather than
stopping individual oscillations \cite{Ukai}.

Our analysis of simple motif combining positive and negative feedback
demonstrates that intrinsic noise, and not just external
perturbations, can also cause switching where a limit cycle and a
stable fixed point coexist in the phase space.  As such repeated
switching would disrupt the circadian rhythm, we can predict that
specific regulatory mechanism must exist in real circadian clocks to
suppress it. It would be interesting to explore the space of small
network motifs to understand what mechanisms could implement this kind
of suppression most effectively.

\section*{Acknowledgement}
We thank Hiroshi Kori and Lei-han Tang for
useful comments. This work is supported by
the Danish National Research Foundation.

\section*{Appendix: The model for Drosophila circadian rhythm}\label{Appendix}

We studied the stochastic version of the model for Drosophila circadian rhythm
given in ref \cite{Leloup2,Leloup3}.
The summary of the reactions in the model with deterministic equations 
of the model are 
shown in Fig. \ref{drosophila} with parameters given in the caption.
We converted these equations into stochastic form
using the Gillespie algorithm \cite{Gillespie}, 
as has been done to convert the simple motifs 
eqs. (\ref{eq1})-(\ref{eq3}) to the stochastic version
(\ref{stoc1})-(\ref{stocN}).
The concentrations are converted to the number of molecules
based on the typical cell volume of drosophila, about $1000 \mu$m$^3$,
which means 1 nM corresponds to about 600 molecules.
Figure \ref{noise_Drosophila}C is based on this conversion.
For the case where the cell volume is 10 (100) fold bigger,
the copy number is also converted to be 10 (100) fold bigger,
which corresponds to the simulations 
shown in Fig. \ref{noise_Drosophila}B (A).

\newpage
\section*{Figure Captions}

\begin{figure}[h]
\begin{center}

\end{center}
\caption{Four patterns of motifs, each consisting a switch and a
  negative feedback loop. Node 1 and node 2 form a switch. Node 1,2,3
  form a negative feedback loop.}
\label{pattern}
\end{figure}

\begin{figure*}[htb]

\caption{The left column is the real 3 dimensional phase space.  The
  middle column is 2 dimensional illustration of the real phase
  space. The right most column shows the right hand side of
  Eq. (\ref{Eq:one_d}). Parameters are set to be 
$h=3$, $B=0.1$, and $A=5$.
(A) $\lambda_3<\beta$, with one stable limit
  cycle (line) and one unstable fixed point.  
(plot with $\beta=2.2$)
(B)
  $\lambda_2<\beta<\lambda_3$, with three fixed points (an unstable
  fixed point with complex eigenvalue (square), saddle node (triangle)
  and a stable fixed point (circle)) and a stable limit cycle.  
(plot with $\beta=2$)
(C)
  $\beta=\lambda_2$, the saddle node hits the limit cycle and forms a
  homoclinic orbit.  
(plot with $\beta=1.923$)
(D) $\lambda_1<\beta<\lambda_2$, there are only
  three fixed points.  
(plot with $\beta=1.2$)
(E) $\beta<\lambda_1$, there is only one stable
  fixed point.  
(plot with $\beta=2$)
}
\label{phase_space}
\end{figure*}

\begin{figure}[htb]
\begin{center}

\end{center}
\caption{
The stable limit cycle  and the stable fixed point 
are shown by solid line and filled circle, respectively,
for $h=3$, $B=0.1$, $A=5$, and $\beta=2$.
The surface shows the boundary between 
the basin of attraction of the stable limit 
cycle and the basin of attraction of the stable fixed point.
}
\label{basin}
\end{figure}

\begin{figure}[htb]
\begin{center}

\end{center}
\caption{Period of oscillations as a function of $\beta$,
with $h=3, B=0.1, A=5$.
The period
  diverges at $\beta=\lambda_2\approx 1.923$ 
(indicated by a vertical line), where a
  homoclinic cycle is present.
As $\beta \to \infty$ the repressor link from node 2 to
node 1 vanishes. What is left is a standard repressilator
and the period one obtains in that limit (equal to 3.7
for these parameter values, shown 
by a horizontal line) is the period of this respressilator.
}
\label{period}
\end{figure}

\begin{figure}
\begin{center}

\end{center}
\caption{
Time evolution of the concentrations 
for the stochastic simulation of the system 
with $h=3, B=0.1, A=5, \beta=2$. All the units are dimensionless, 
and concentrations are converted to the numbers 
so that $X_i=1$ corresponds to one molecule when $V=1$.
%$k=1, \gamma=1, h=3, c=0.1, \alpha=1, \beta=2$, 
(A) $V=1000$  and (B) $V=100$.
}
\label{noise_3_node}
\end{figure}

\begin{figure}
\begin{center}

\end{center}
\caption{
Switching rates from the oscillatory state to the 
steady state (open circles)
and from the steady state to the oscillatory state (filled circles),
as a function of system volume $V$, for the same parameter values 
as in Fig. \ref{noise_3_node}.
%which is proportional to the copy number of molecules.
The threshold values are set to be $S_1=0.316$,  
$S_2=3.317$.
}
\label{switchrate}
\end{figure}

\begin{figure}

\caption{Intrinsic noise is introduced to the original model of
  Drosophila \cite{Leloup1,Leloup2}. 
Similar phenomena are observed. When cell size is very
  large, namely about 100 times of a typical cell volume, i.e. $10^5
  \mu m$ (A), the system would switch from oscillation state to steady
  state at a random moment due to noise.  As the cell size goes down
  ($10^4 \mu m$ in B, $10^3 \mu m$ in C), the switching becomes more
  and more frequent.  A typical cell volumes for Drosophila is about
  $10^3 \mu m$ (C), which exhibit very noisy dynamics.  
Detailed description of the model 
with the parameter set is given in appendix, Fig. \ref{drosophila}.
}
\label{noise_Drosophila}
\end{figure}

\begin{figure*}

\caption{
The model for Drosophila circadian rhythm
given in ref \cite{Leloup2,Leloup3}. 
Negative feedbacks are shown with dotted line, while 
positive feedback loops are shown with dashed line.
The deterministic equation of the model is shown in the 
right panel, where variables are concentrations of 
{\sl per} ($M_p$) and tim ($M_T$) mRNAs,
the PER and TIM with three phosphorylation levels of $P_0$ ($T_0$),
$P_1$ ($T_1$), and $P_2$ ($T_2$), respectively, 
the PER-TIM complex $C$, and the nuclear form of the PER-TIM complex
($C_N$).
Parameters used are:
$n = 4$,  $v_{sP}$ = 1.1 nM/h,  $v_{sT}$ = 1 nM/h,  $v_{mP}$ =1.0 nM/h, 
$v_{mT}$ = 0.7nM/h, $v_{dP}$ = 2.2nM/h, $k_{sP} = k_{sT}$ = 0.9 /h, 
$k_1$ = 0.8 /h, $k_2$ = 0.2 /h, $k_{3}$ = 1.2/(nM$\cdot$ h), $k_4$ = 0.6 /h, 
$K_{mP} =K_{mT}$ = 0.2 nM, $K_{IP} =K_{IT} $=1nM, 
$K_{dP} =K_{dT}$=0.2nM, 
$K_{1P}=K_{1T}=K_{2P}=K_{2T}=K_{3P}=K_{3T}=K_{4P}=K_{4T}$=2nM, 
$V_{1P}=V_{1T}$=8nM/h, 
$V_{2P}=V_{2T}$=1nM/h, $V_{3P}=V_{3T}$=8nM/h, 
$V_{4P}=V_{4T}$= 1 nM/h, 
$k_d = k_{dC} = k_{dN} $= 0.01 /h, 
$V_{dT}$= 1.3nM/h.
}\label{drosophila}
\end{figure*}

\clearpage
\newpage
Figure 1

%\begin{figure}[h]
\begin{center}
\includegraphics[width=0.20\textwidth]{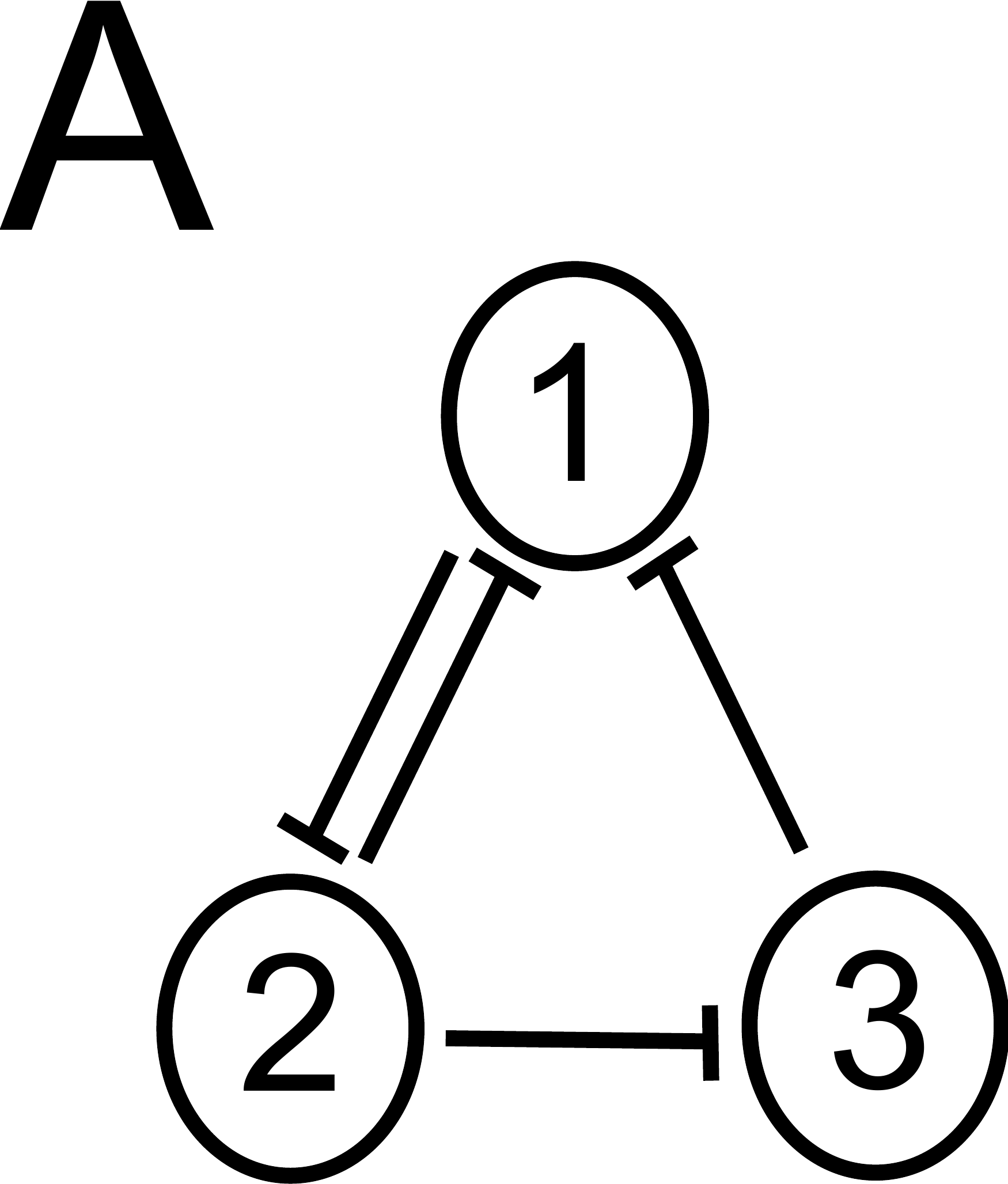}
\includegraphics[width=0.20\textwidth]{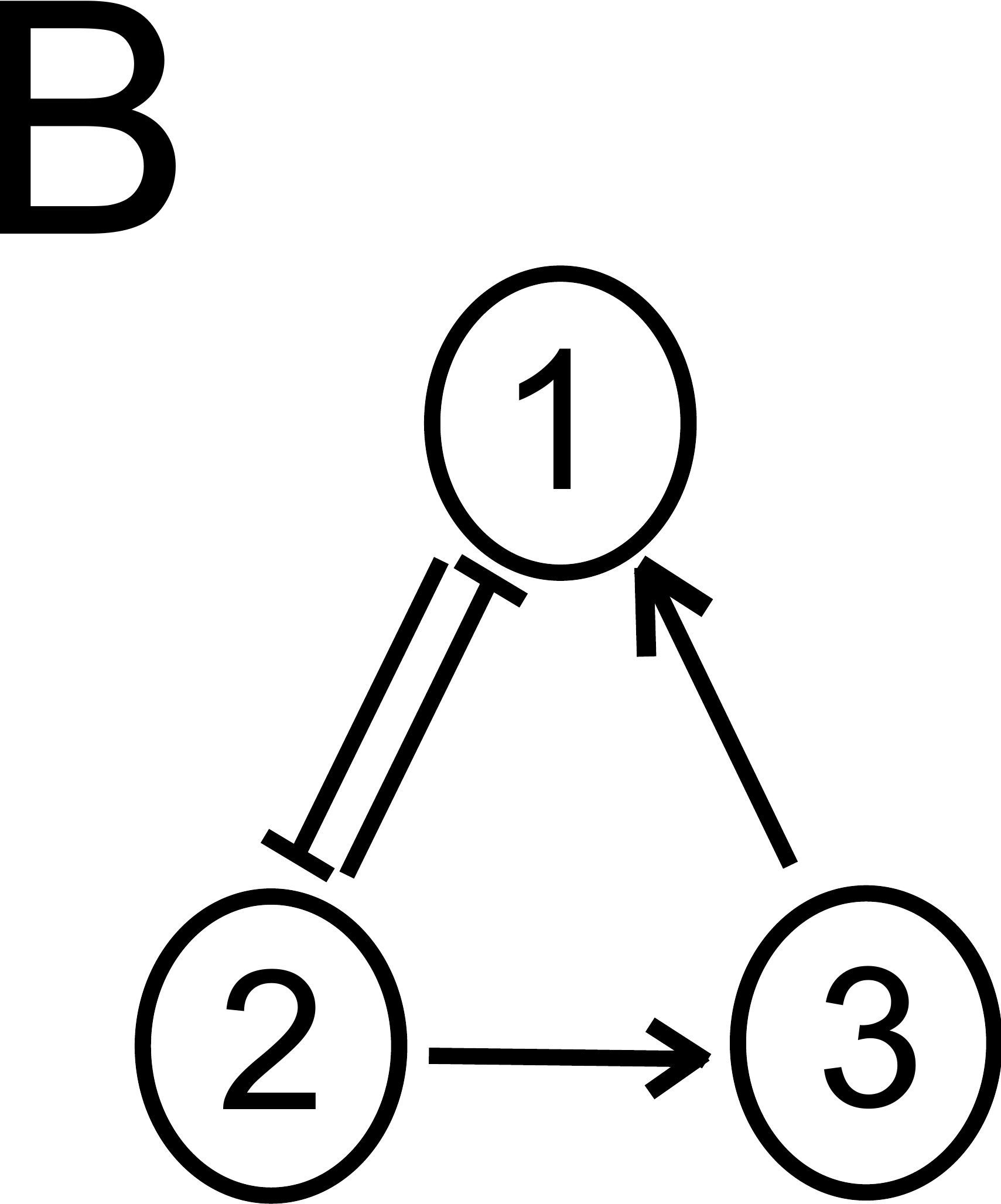}
\includegraphics[width=0.20\textwidth]{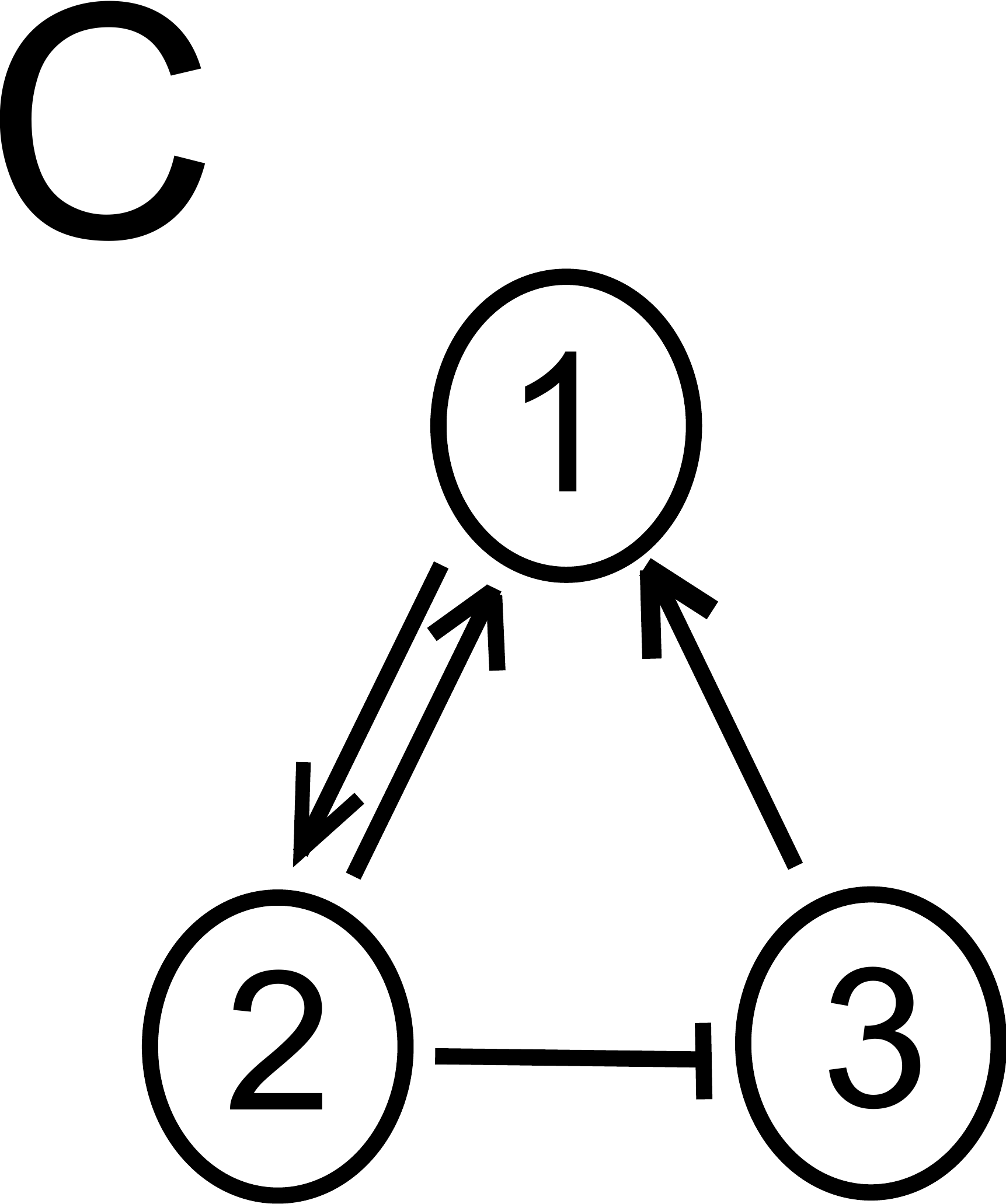}
\includegraphics[width=0.20\textwidth]{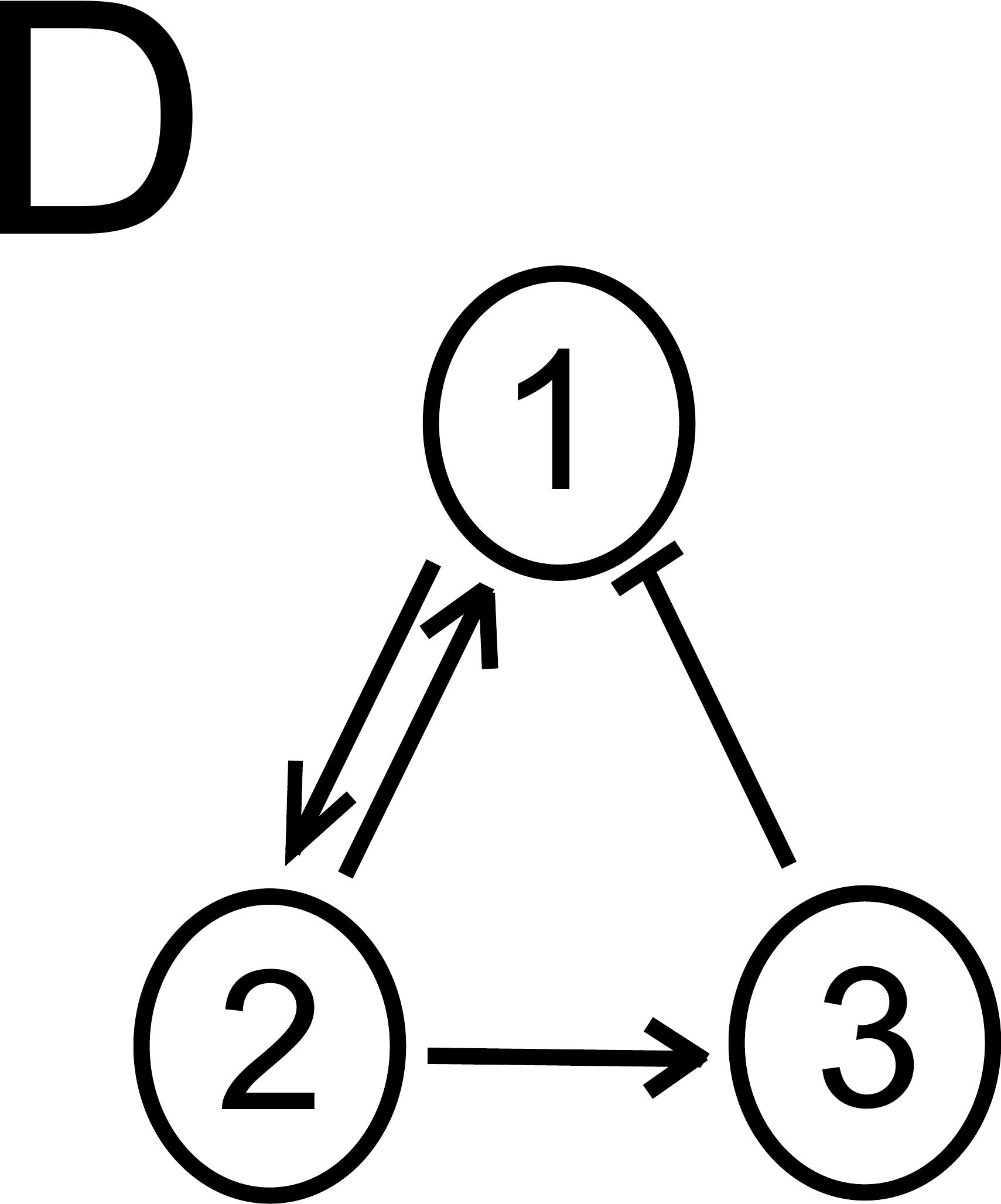}
\end{center}
%\caption{Four patterns of motifs, each consisting a switch and a
%  negative feedback loop. Node 1 and node 2 form a switch. Node 1,2,3
%  form a negative feedback loop.}
%\label{pattern}
%\end{figure}

\newpage
Figure 2

%\begin{figure*}[htb]
\begin{center}
\includegraphics[width=0.25\textwidth]{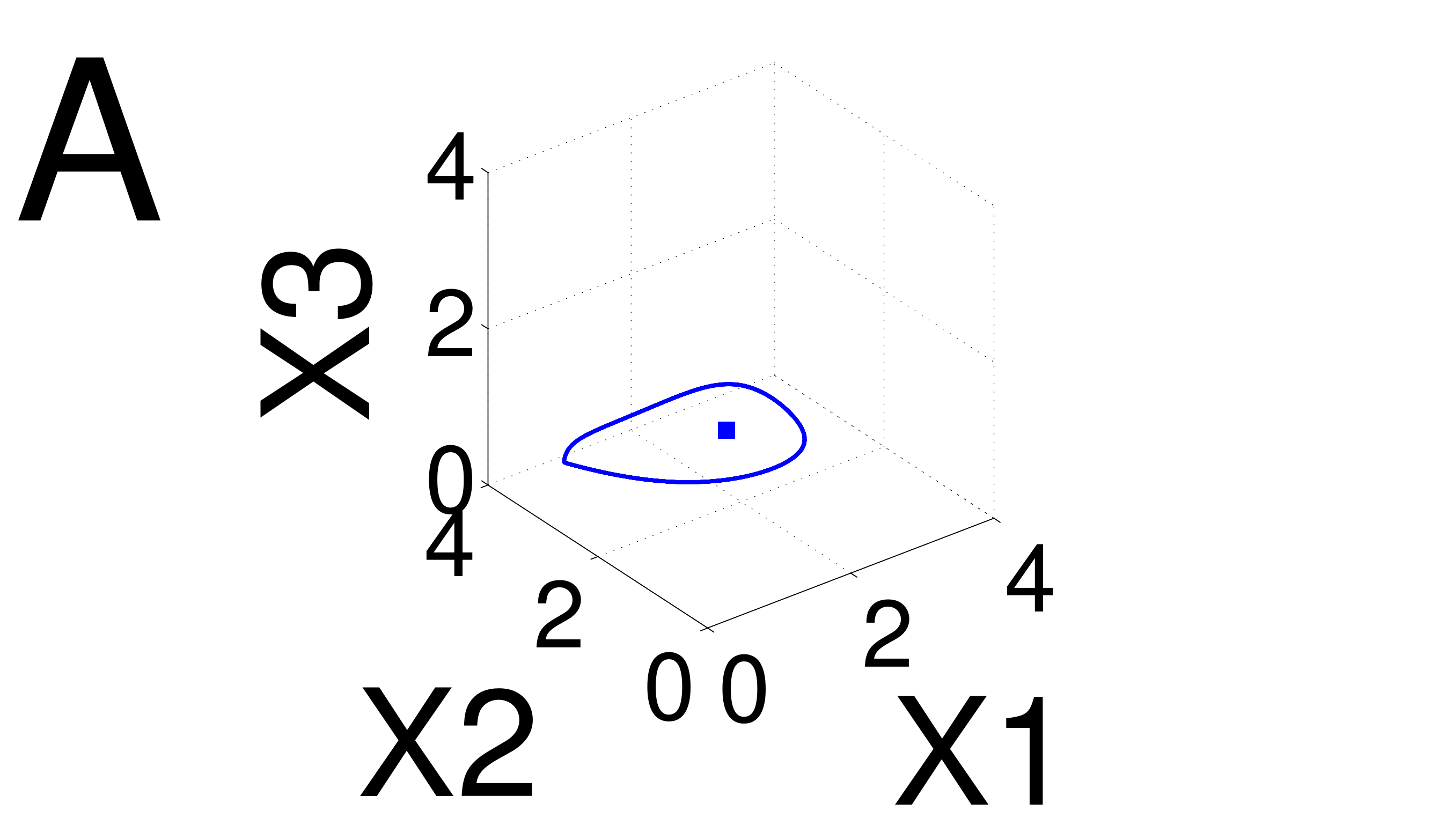}
\includegraphics[width=0.1\textwidth]{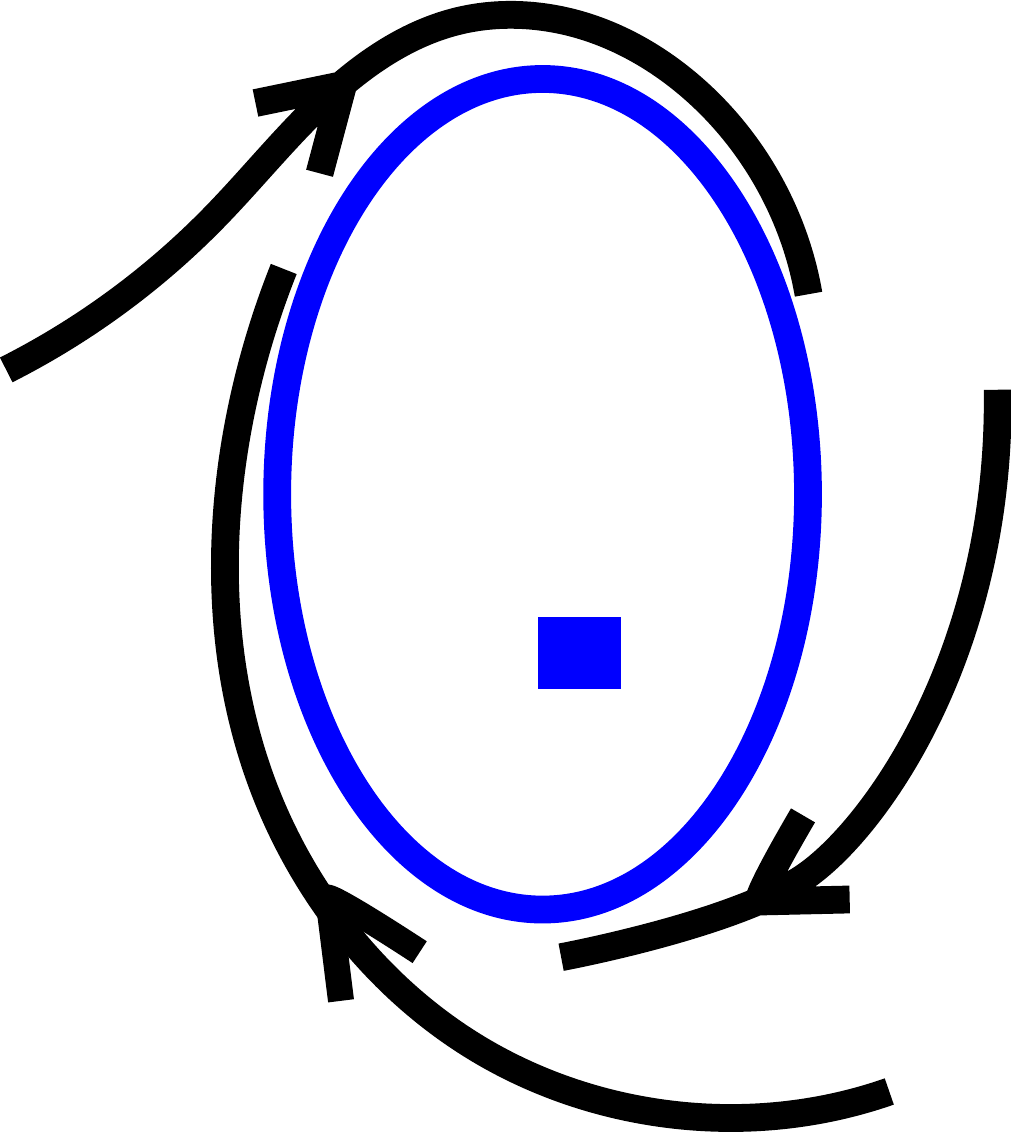}
\includegraphics[width=0.20\textwidth]{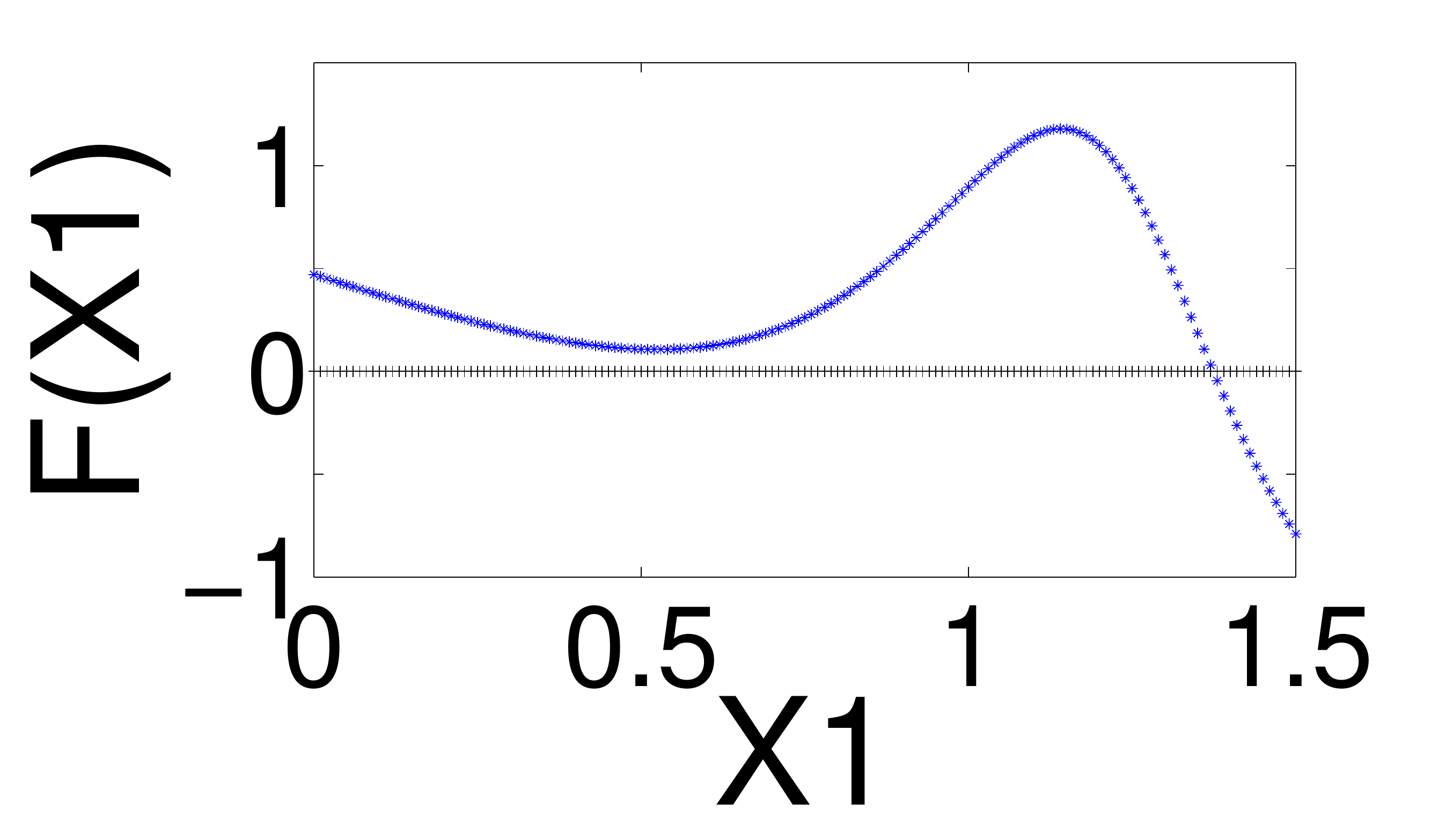}\\
\includegraphics[width=0.25\textwidth]{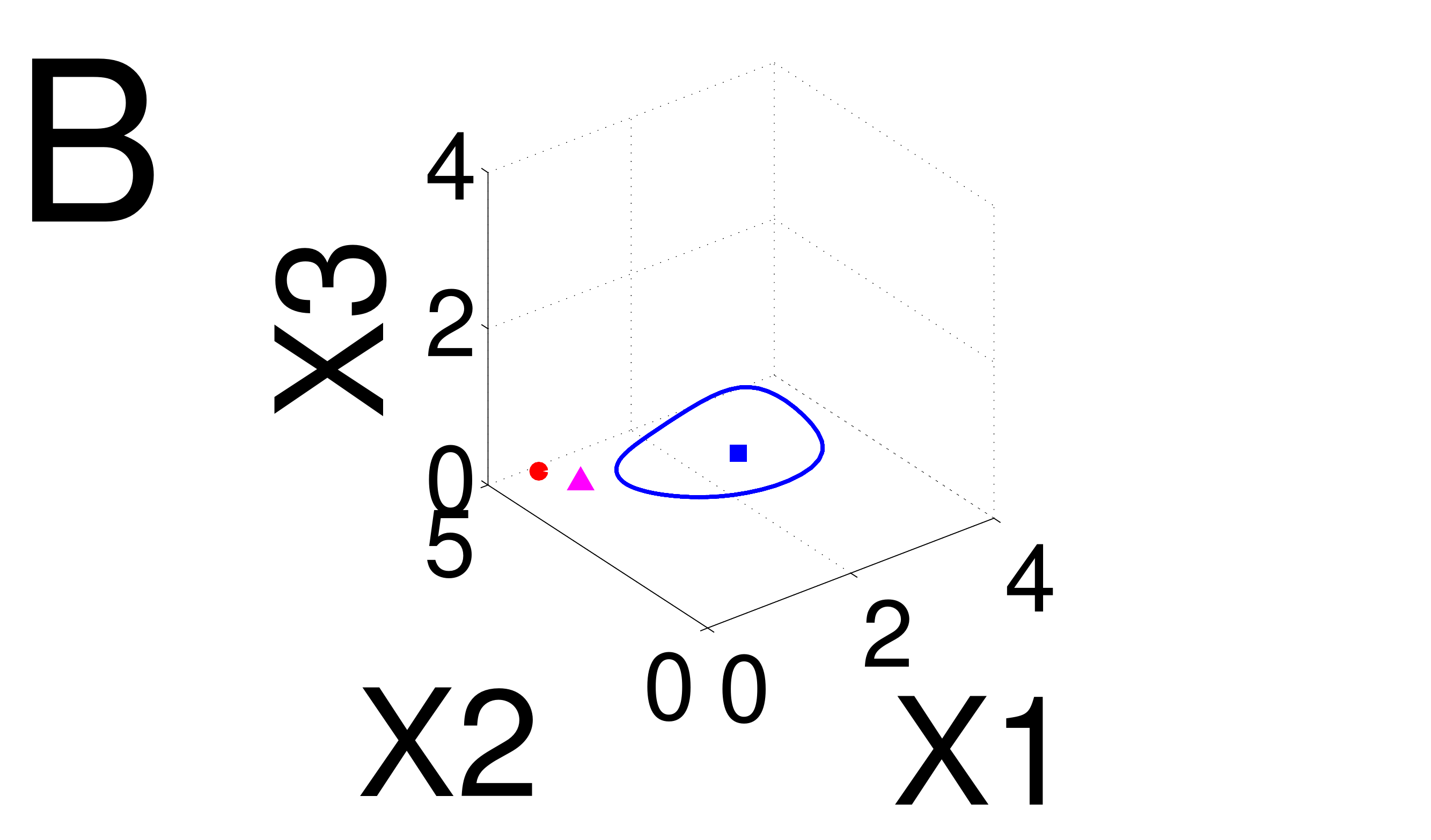}
\includegraphics[width=0.1\textwidth]{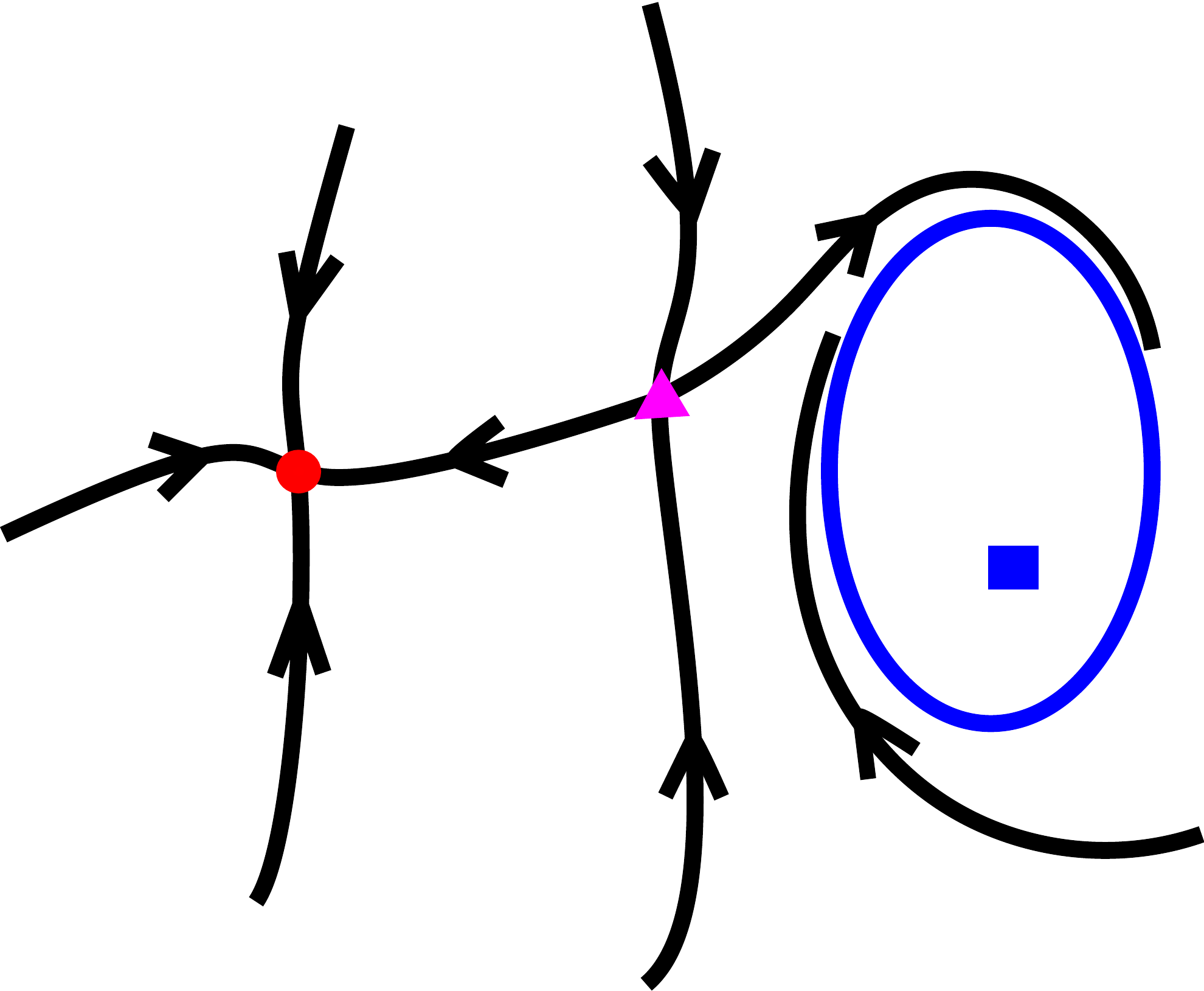}
\includegraphics[width=0.2\textwidth]{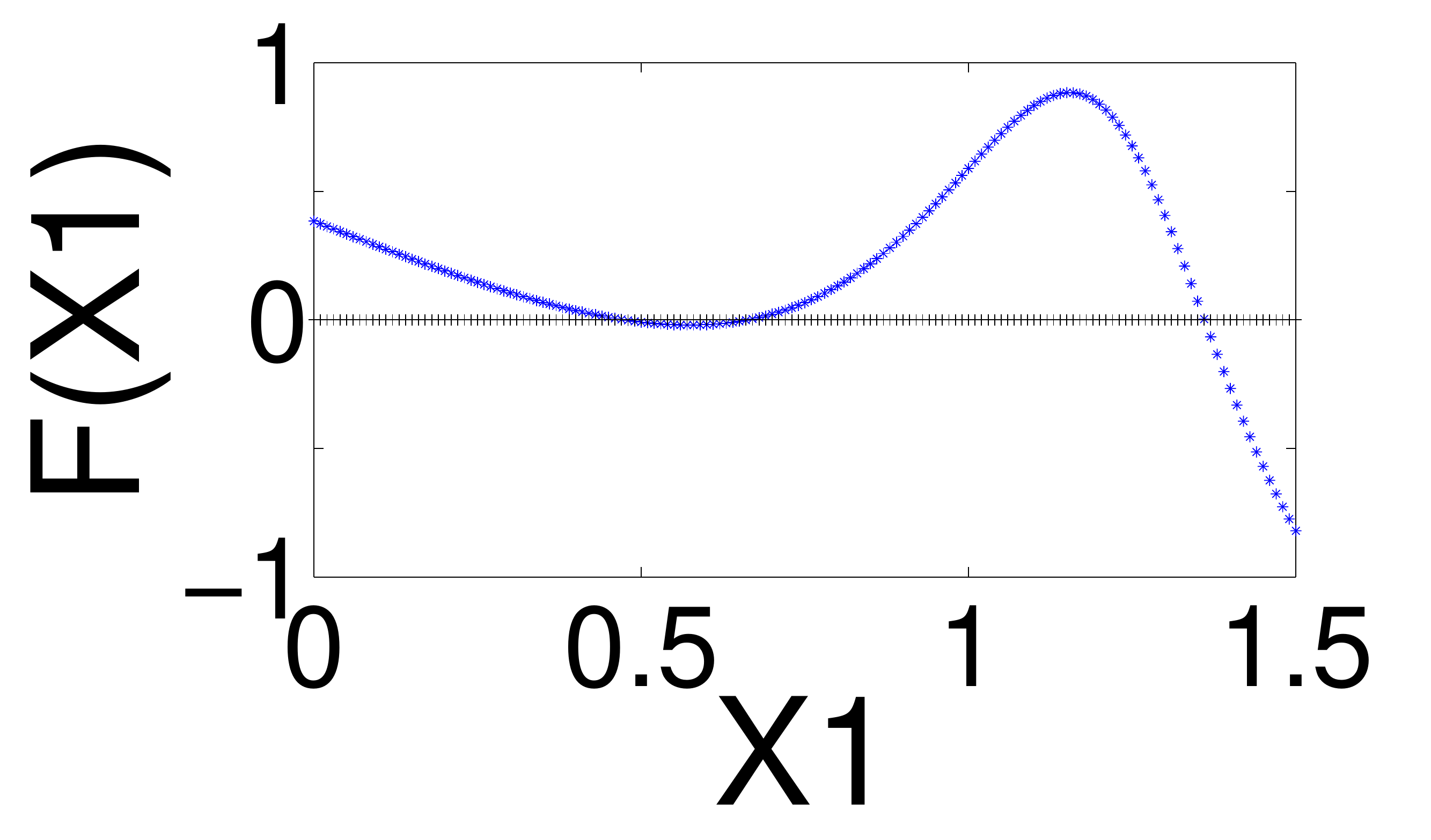}\\
\includegraphics[width=0.25\textwidth]{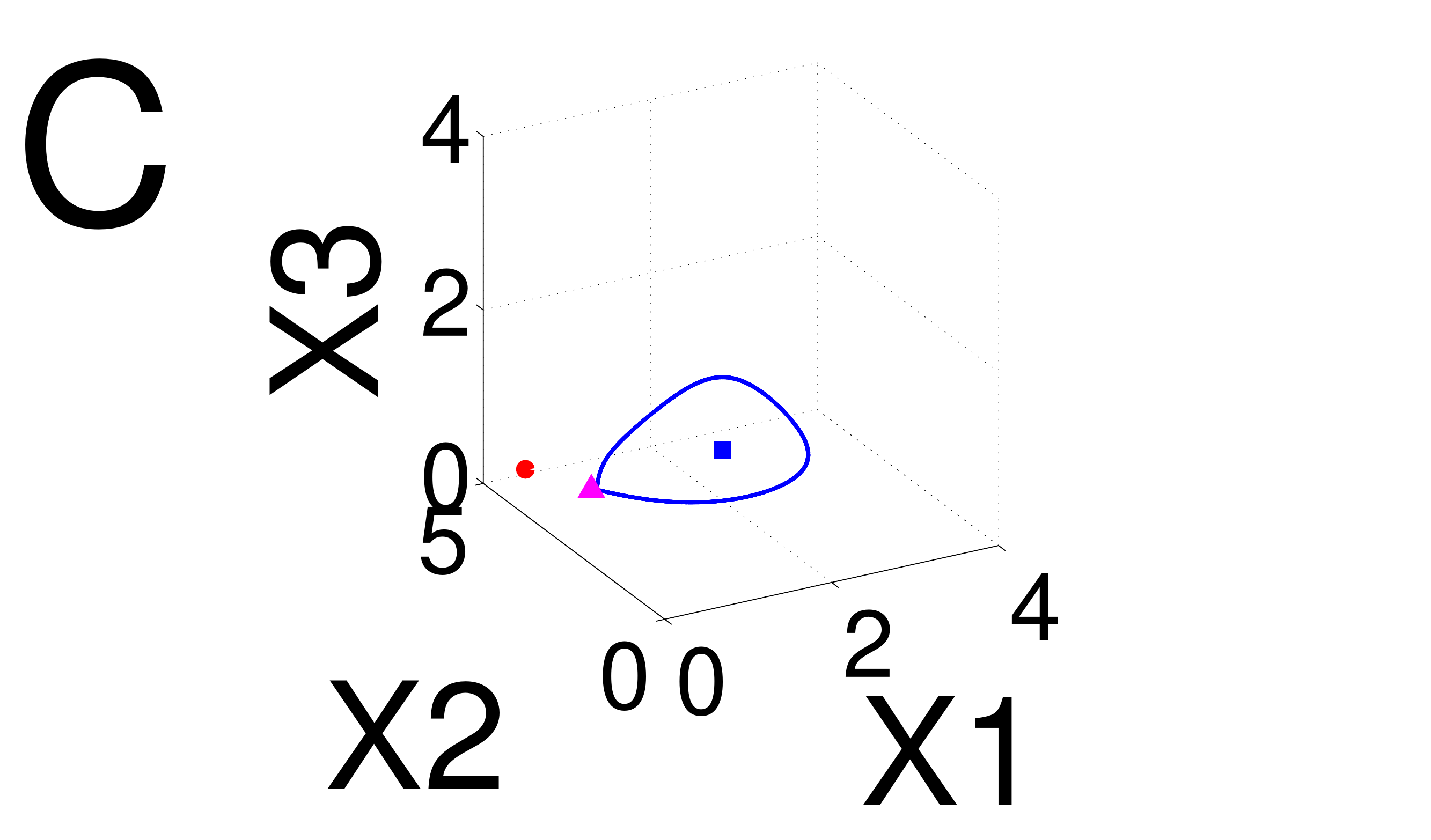}
\includegraphics[width=0.1\textwidth]{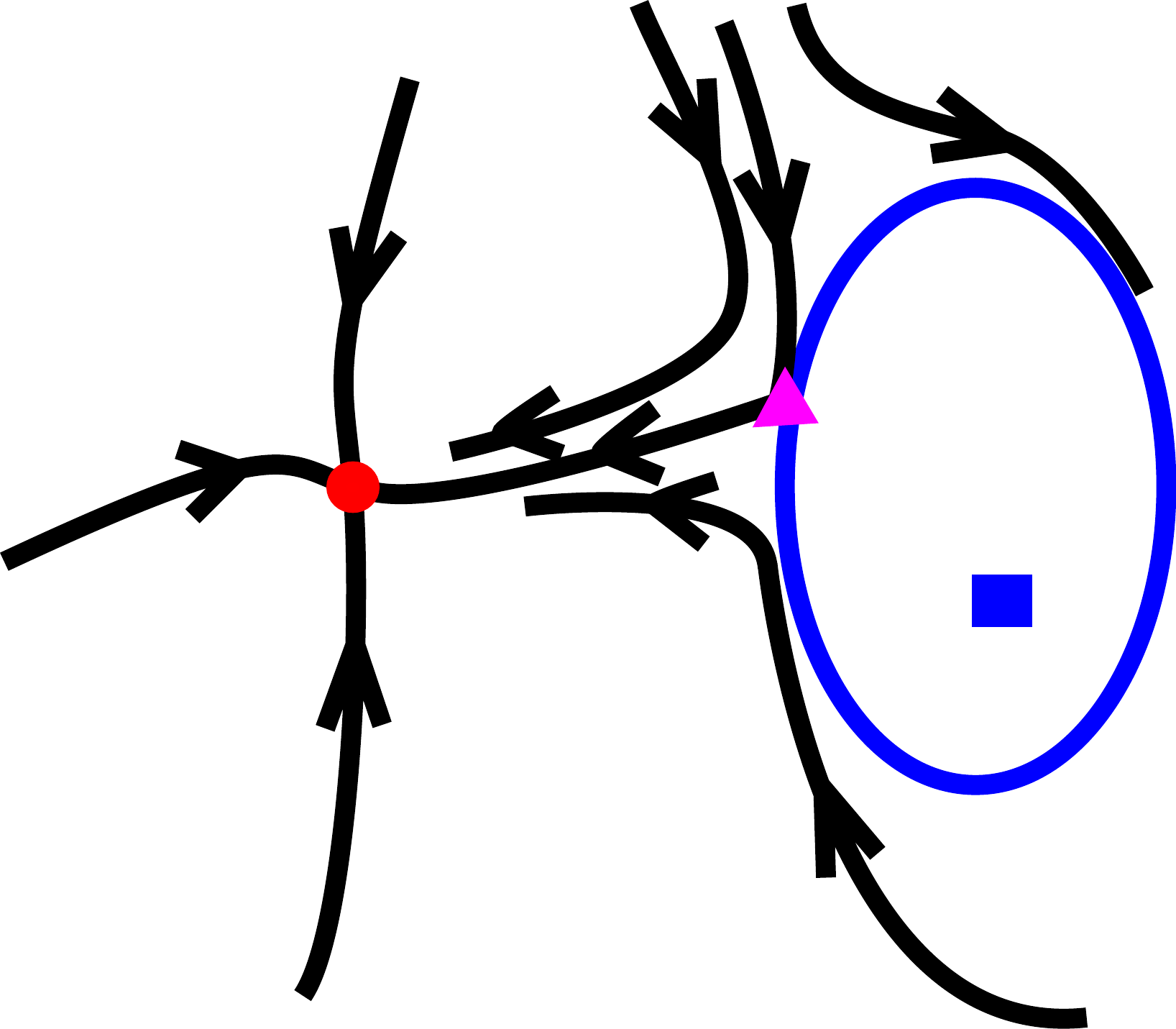}
\includegraphics[width=0.20\textwidth]{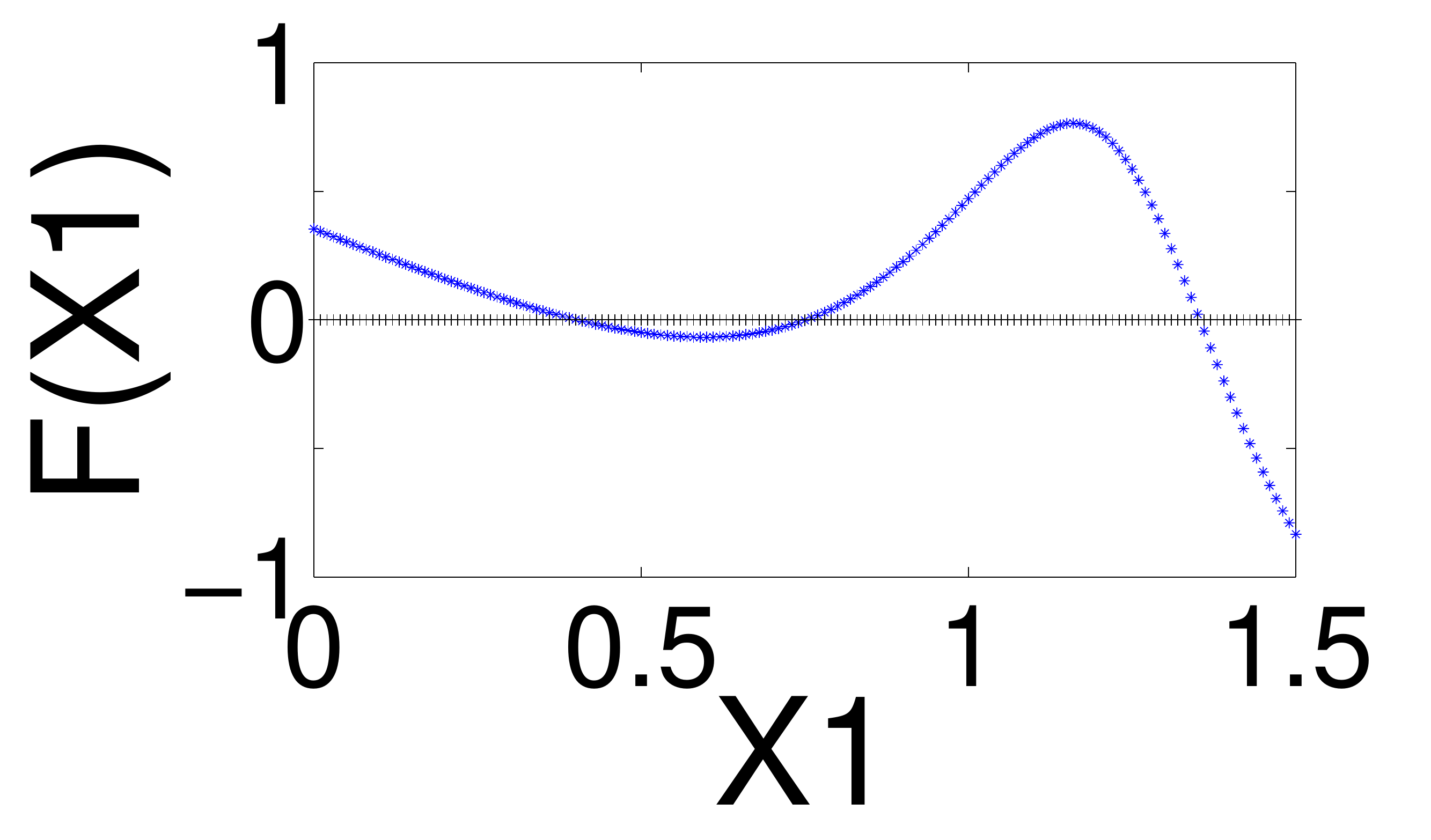}\\
\includegraphics[width=0.25\textwidth]{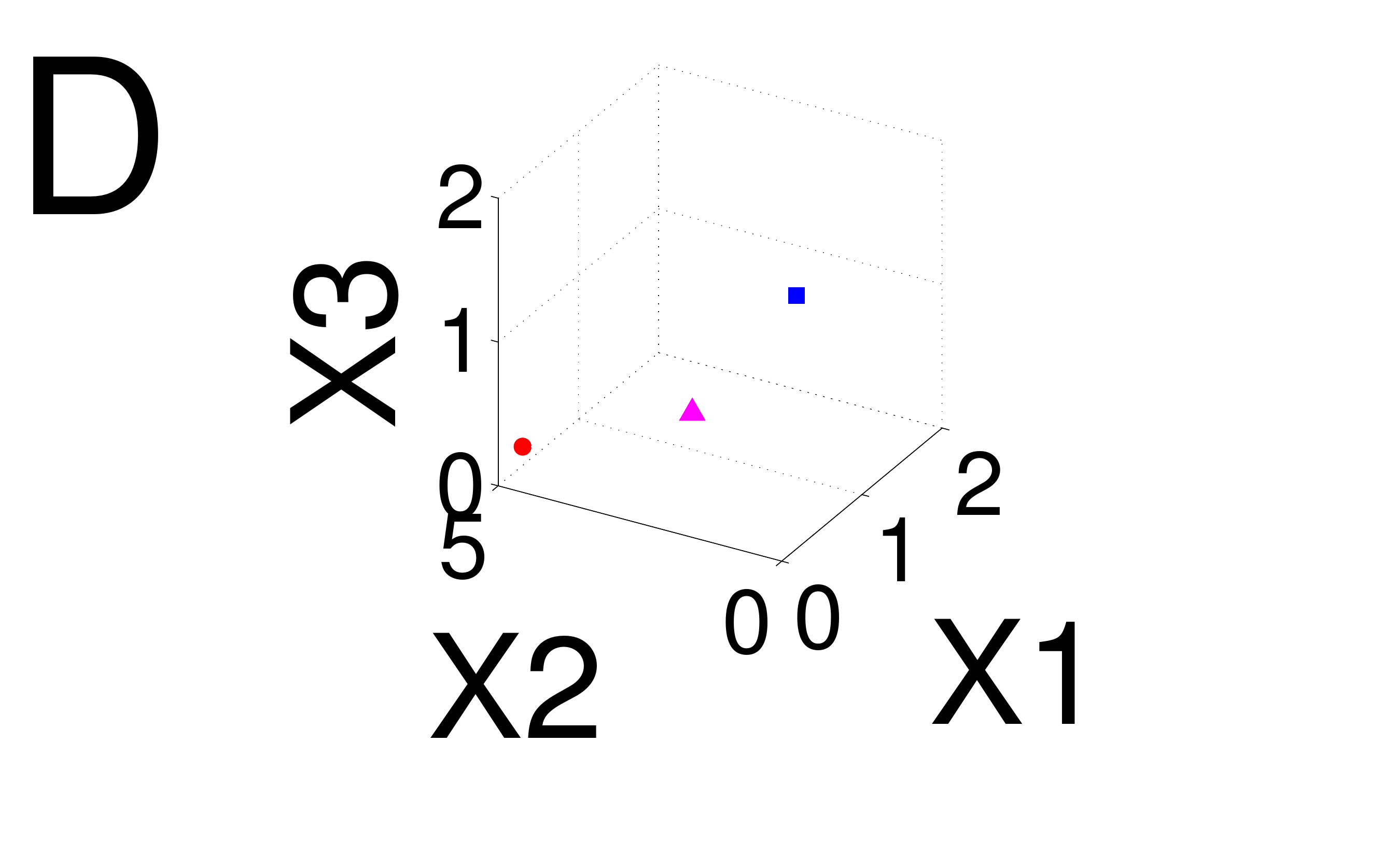}
\includegraphics[width=0.1\textwidth]{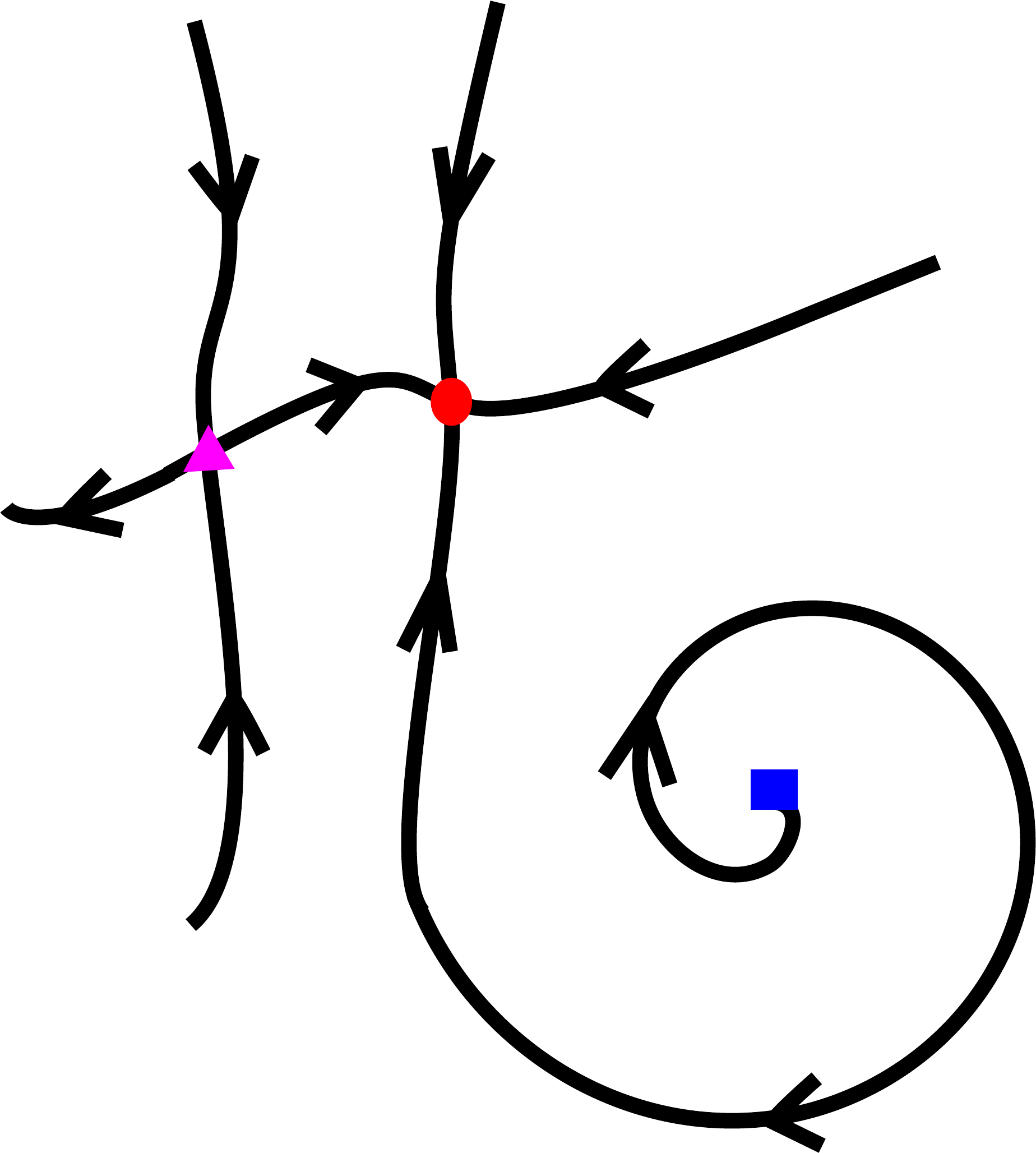}
\includegraphics[width=0.20\textwidth]{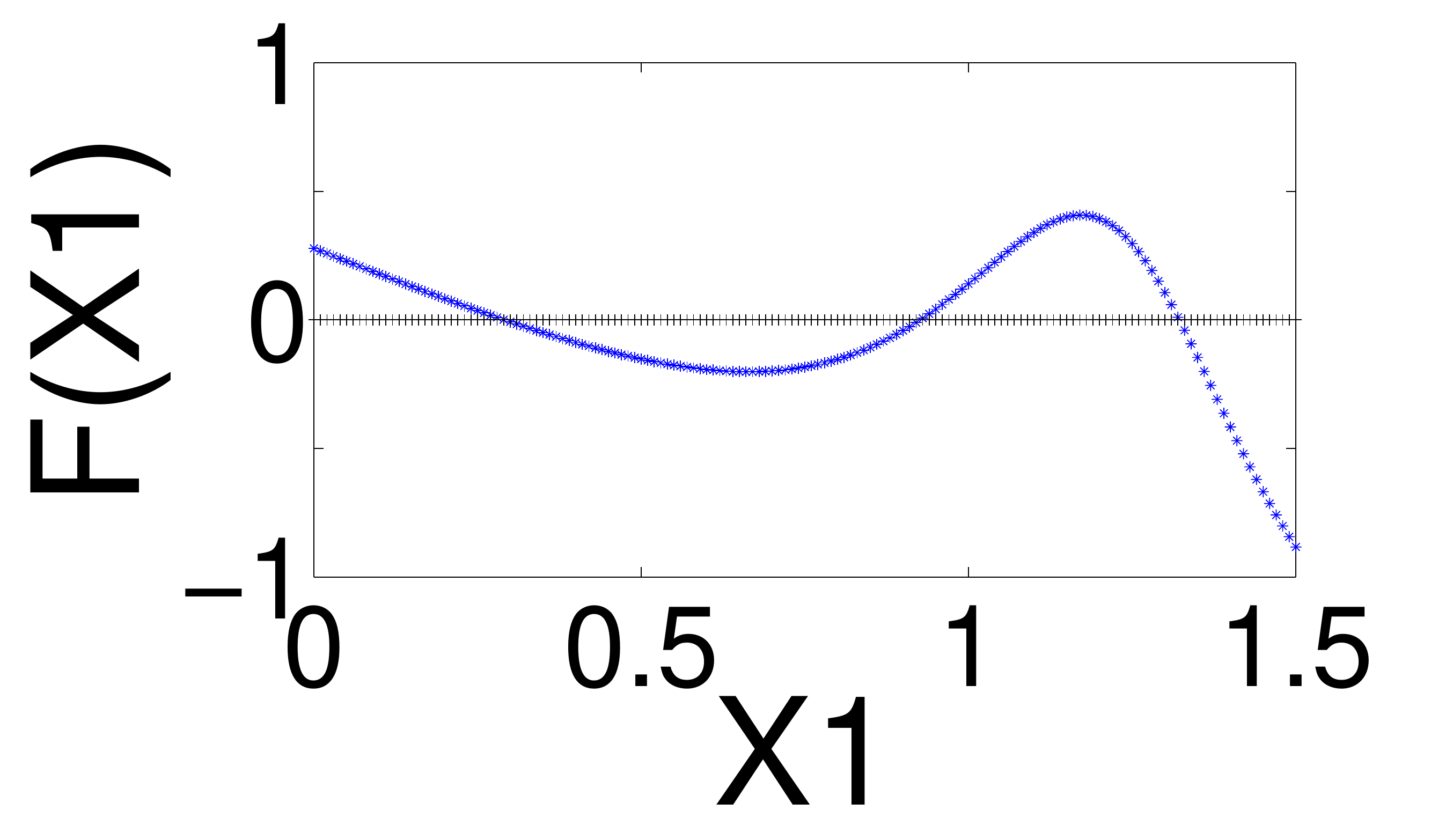}\\
\includegraphics[width=0.25\textwidth]{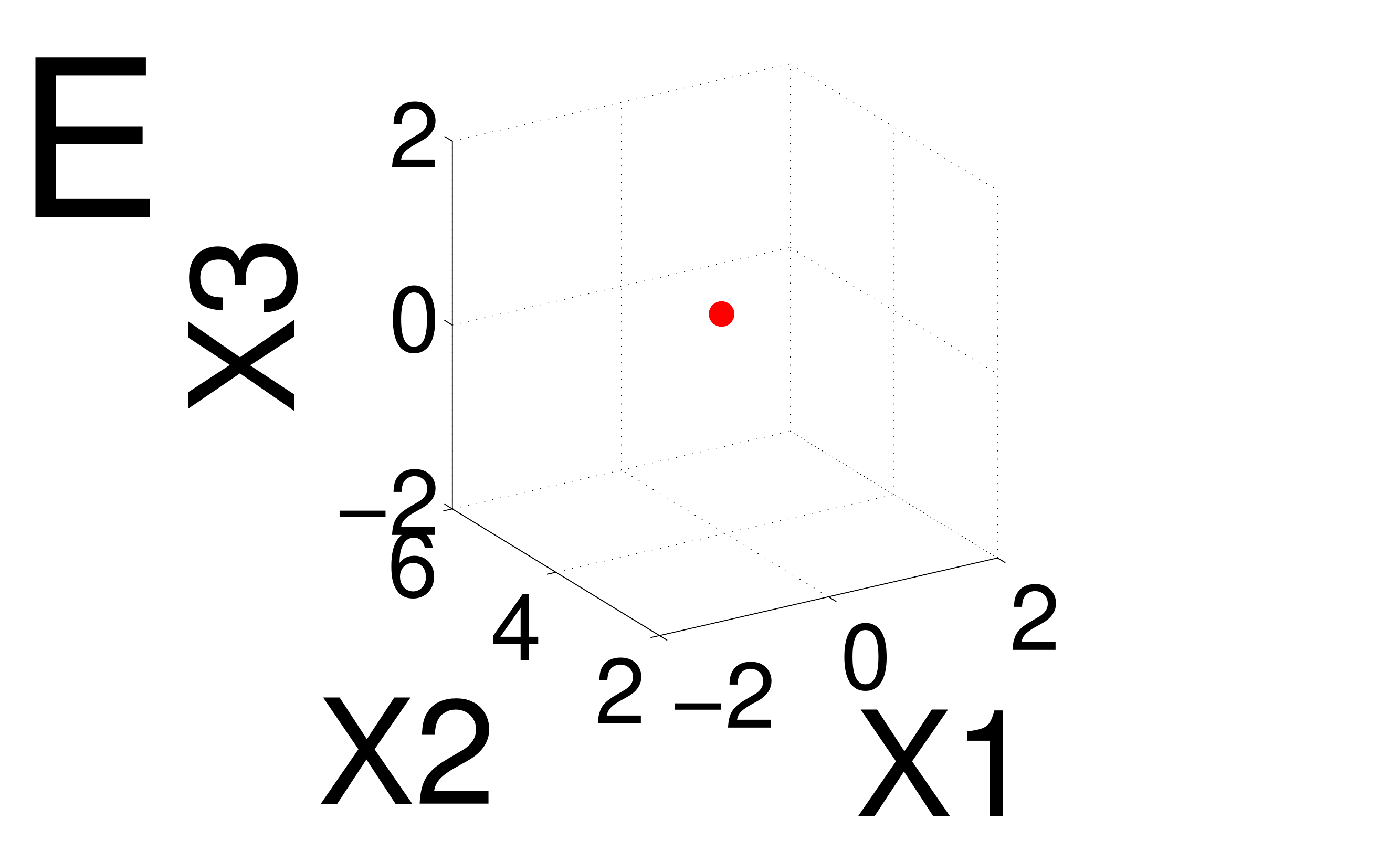}
\includegraphics[width=0.1\textwidth]{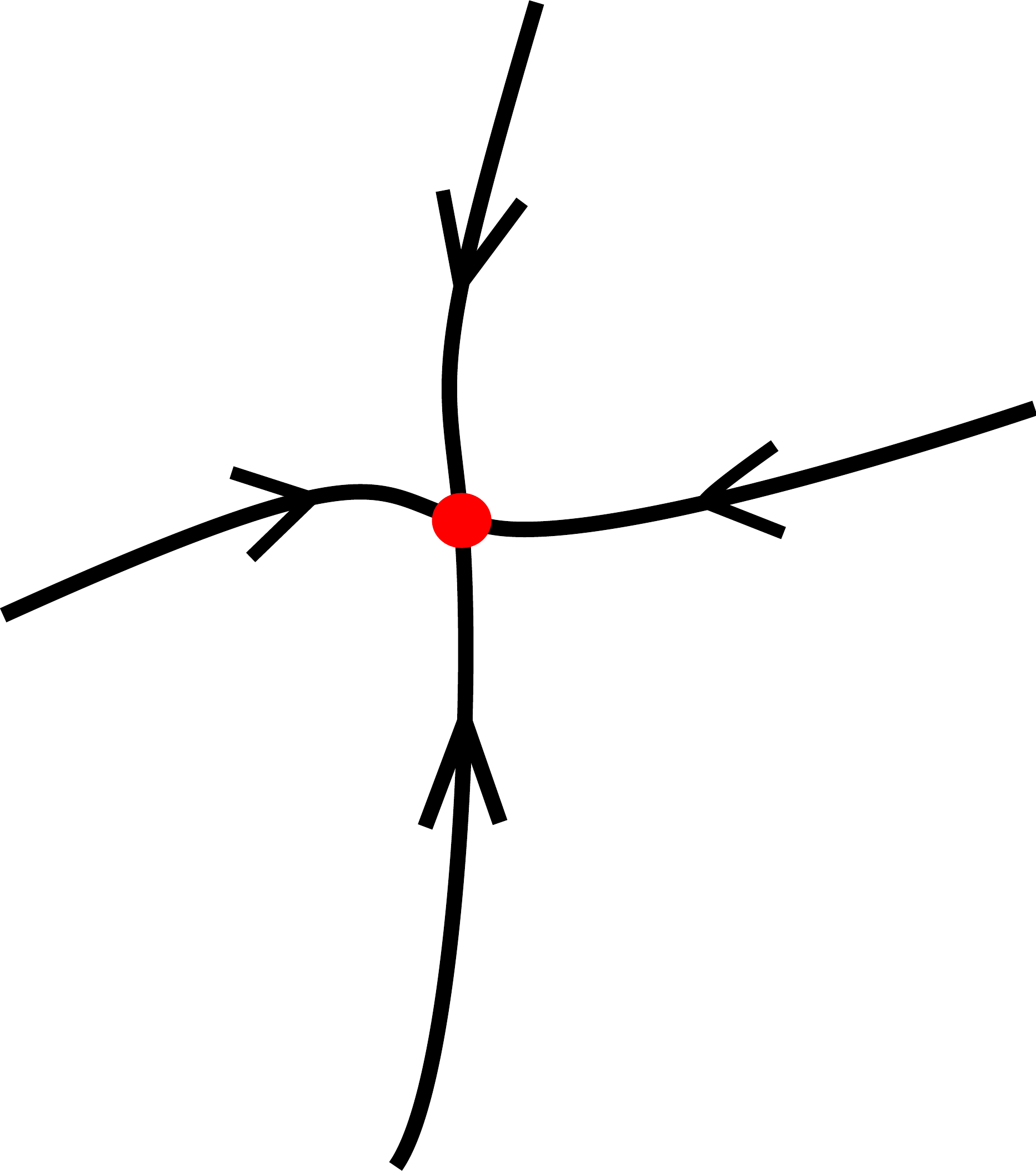}
\includegraphics[width=0.20\textwidth]{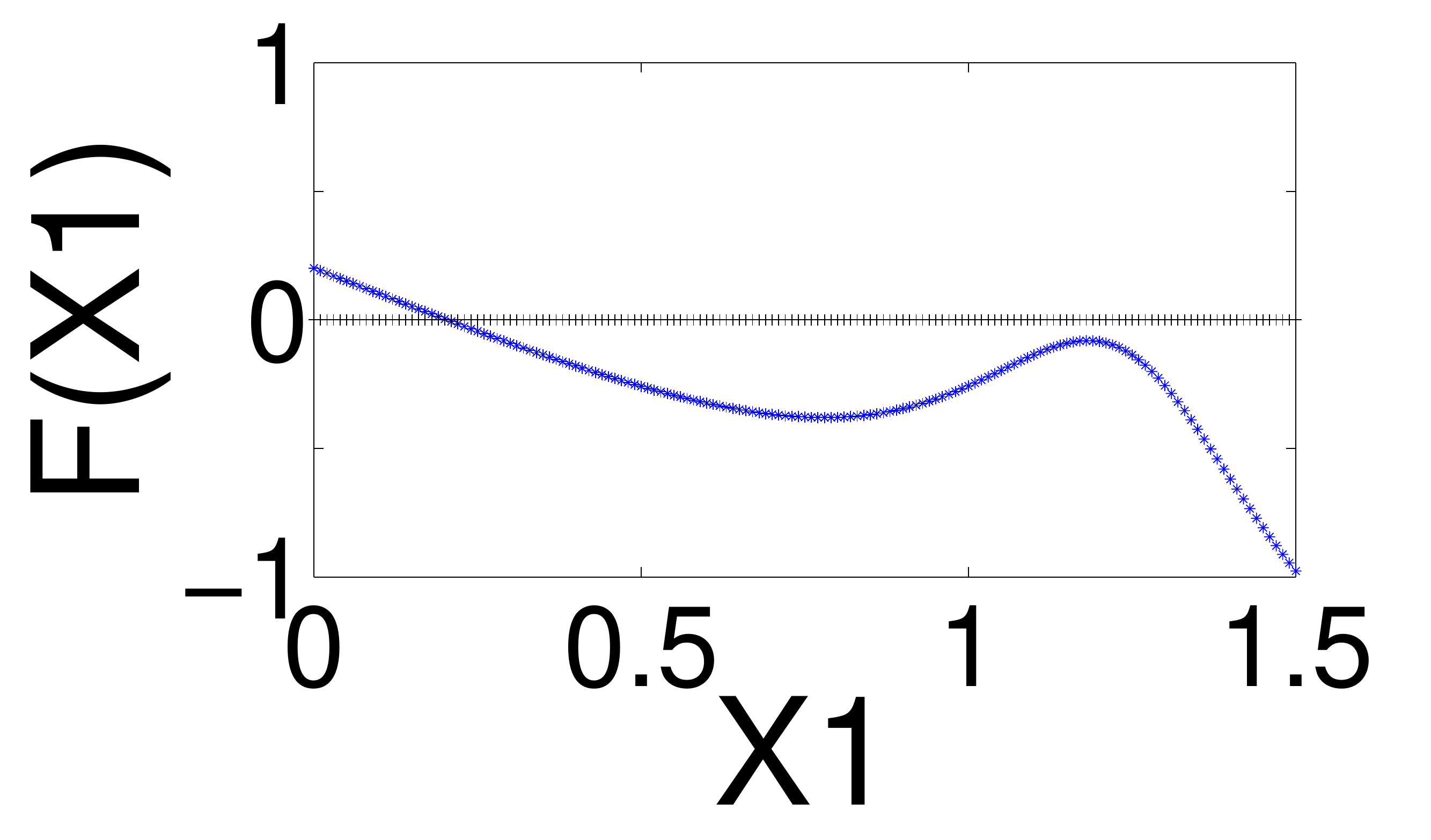}
\end{center}
%\caption{The left column is the real 3 dimensional phase space.  The
%  middle column is 2 dimensional illustration of the real phase
%  space. The right most column shows the right hand side of
%  Eq. (\ref{Eq:one_d}).  (A) $\lambda_3<\beta$, with one stable limit
%  cycle (blue line) and one unstable fixed point.  (B)
%  $\lambda_2<\beta<\lambda_3$, with three fixed points (an unstable
%  fixed point with complex eigenvalue (square), saddle node (triangle)
%  and a stable fixed point (circle)) and a stable limit cycle.  (C)
%  $\beta=\lambda_2$, the saddle node hits the limit cycle and forms a
%  homoclinic orbit.  (D) $\lambda_1<\beta<\lambda_2$, there are only
%  three fixed points.  (E) $\beta<\lambda_1$, there is only one stable
%  fixed point.  }
%\label{phase_space}
%\end{figure*}
\newpage
Figure 3
%\begin{figure}[htb]
\begin{center}
\includegraphics[width=0.5\textwidth]{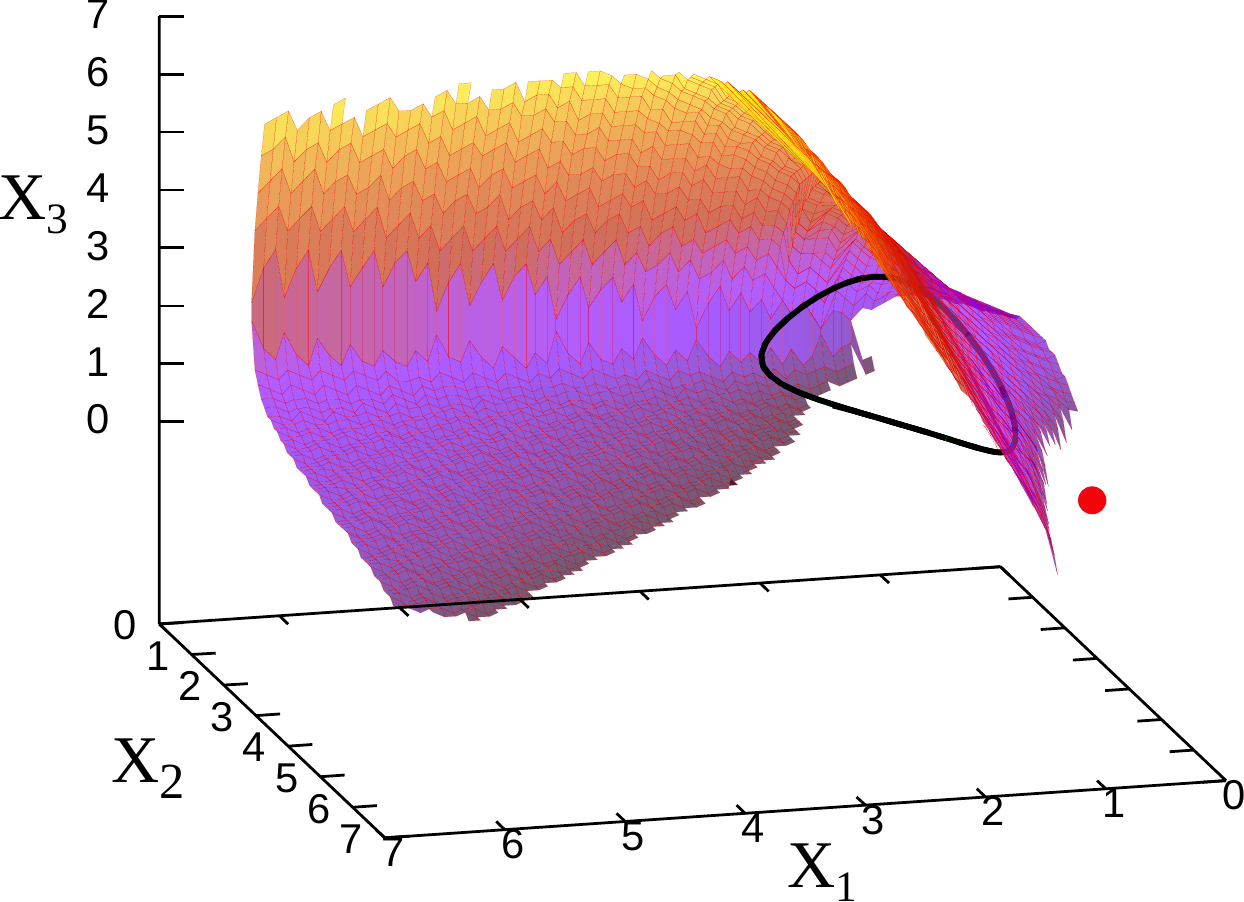}
\end{center}
%\caption{
%The stable limit cycle  and the the stable fixed point 
%are shown by solid line and filled circle, respectively.
%The surface shows the boundary between 
%the basin of attraction of the stable limit 
%cycle and the basin of attraction of the stable fixed point.
%}
%\label{basin}
%\end{figure}

\newpage
Figure 4
%\begin{figure}[htb]
\begin{center}
\includegraphics[width=0.5\textwidth]{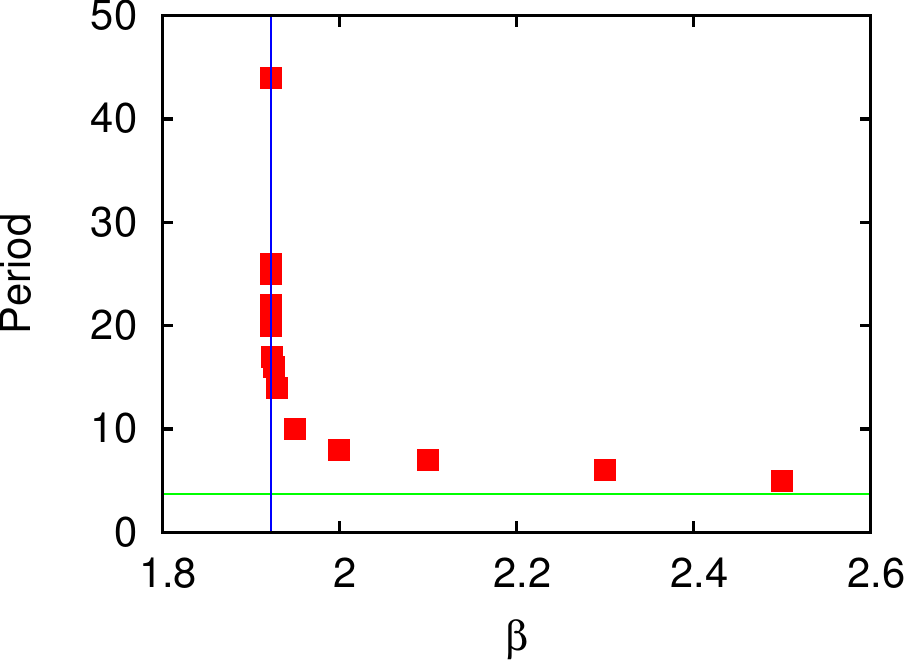}
\end{center}
%\caption{Period of oscillations as a function of $\beta$. The period
%  diverges at $\beta=\lambda_2\approx 1.923$, where a
%  homoclinic cycle is present.}
%\label{period}
%\end{figure}
\newpage
Figure 5
%\begin{figure}
\begin{center}
\includegraphics[width=0.5\textwidth]{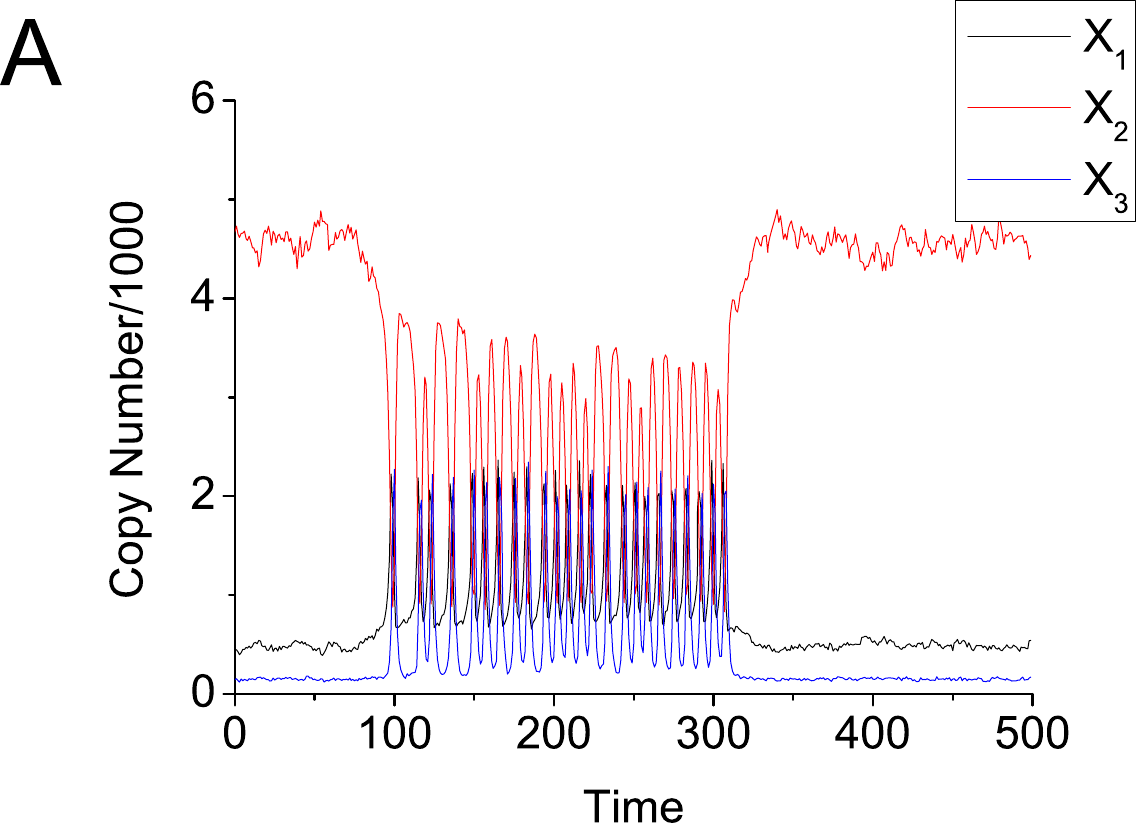}
\includegraphics[width=0.5\textwidth]{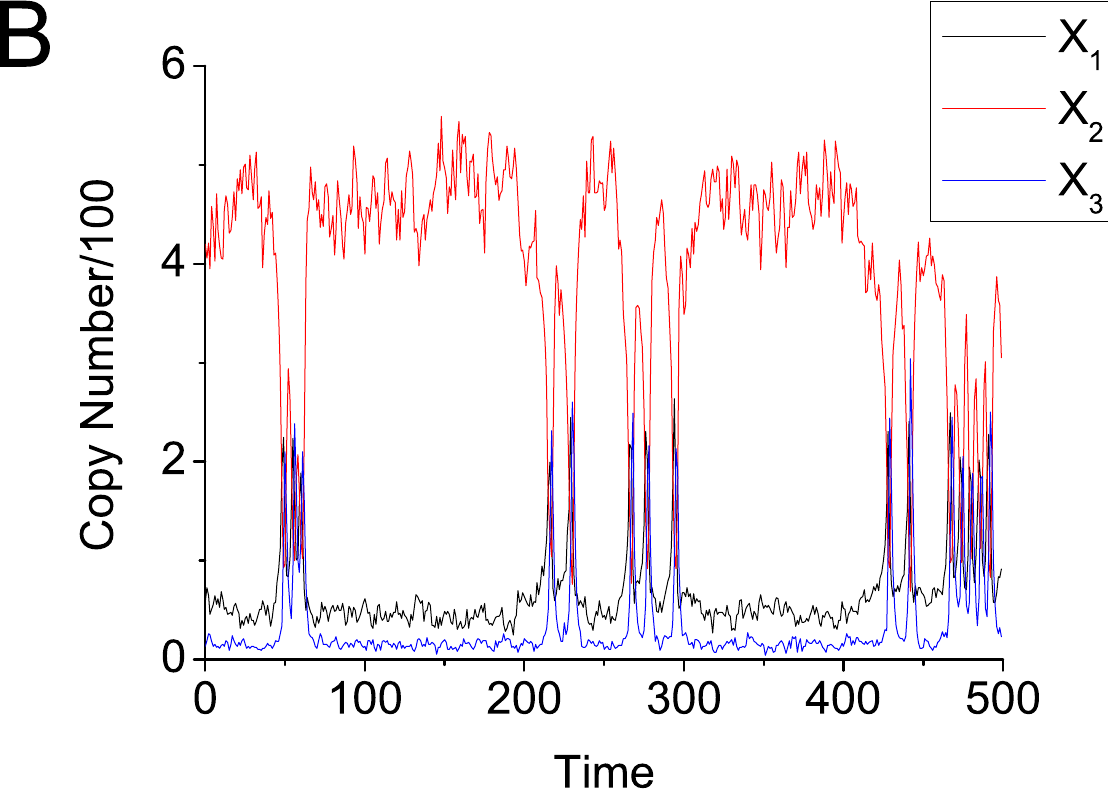}
\end{center}
%\caption{
%Time evolution of the concentrations 
%for the stochastic simulation of the system 
%with $h=3, B=0.1, A=1, \beta=2$. All the units are dimensionless, 
%and concentrations are converted to the numbers 
%so that $X_i=1$ corresponds to one molecule when $V=1$.
%%$k=1, \gamma=1, h=3, c=0.1, \alpha=1, \beta=2$, 
%(A) $V=1000$  and (B) $V=100$.
%}
%\label{noise_3_node}
%\end{figure}

\newpage
Figure 6
%\begin{figure}
\begin{center}
\includegraphics[width=0.5\textwidth]{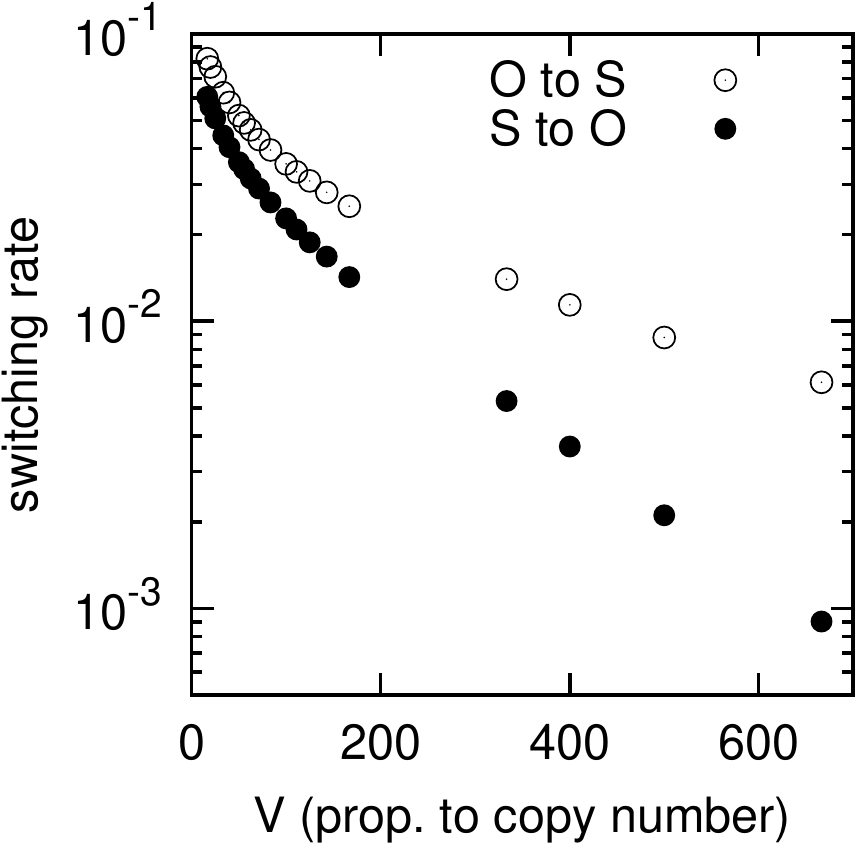}
\end{center}
%\caption{
%Switching rates from the oscillatory state to the 
%steady state (open circles)
%and from the steady state to the oscillatory state (filled circles),
%as a function of system volume $V$,
%which is proportional to the copy number of molecules.
%}
%\label{switchrate}
%\end{figure}
%
\newpage
Figure 7
%\begin{figure}
\begin{center}
\includegraphics[width=0.5\textwidth]{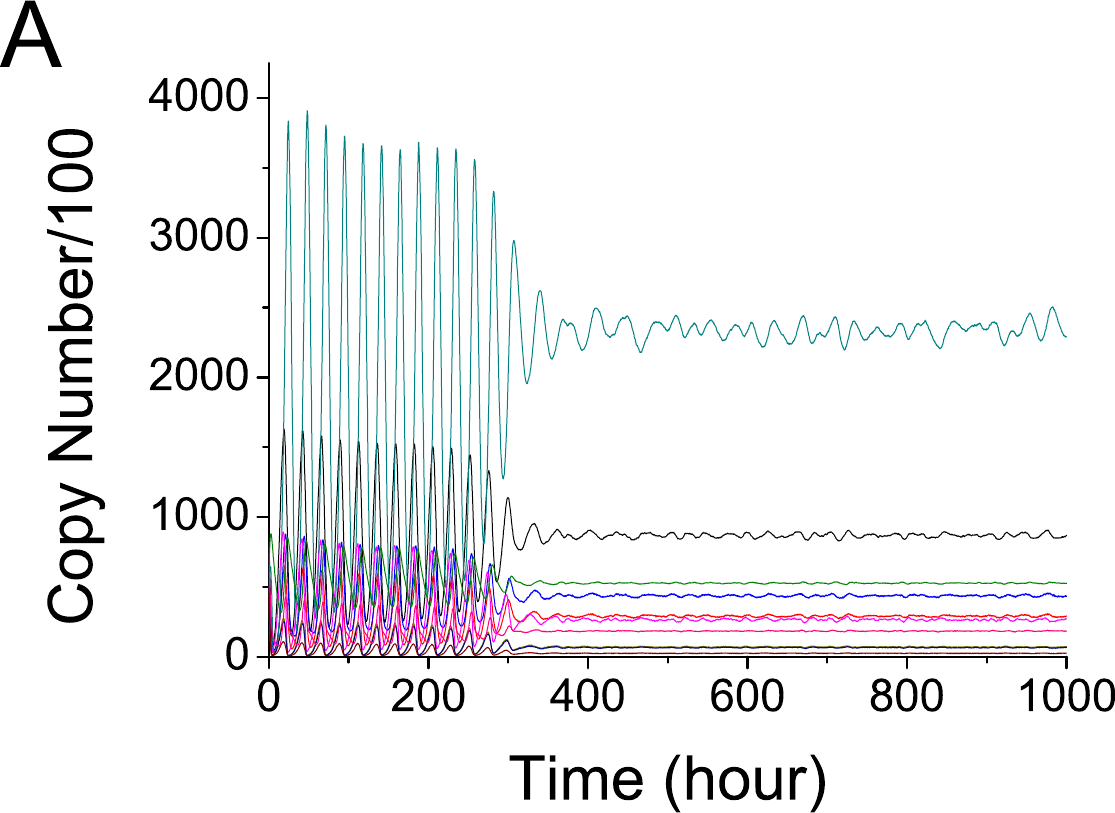}
\includegraphics[width=0.5\textwidth]{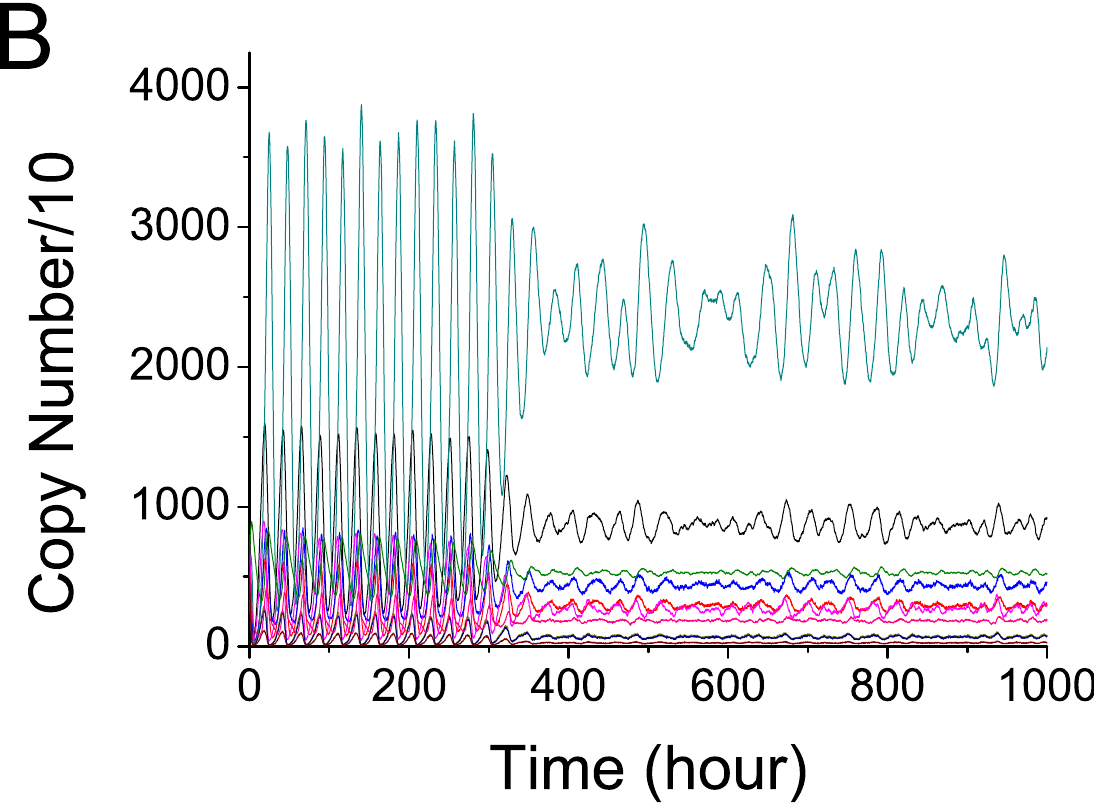}
\includegraphics[width=0.5\textwidth]{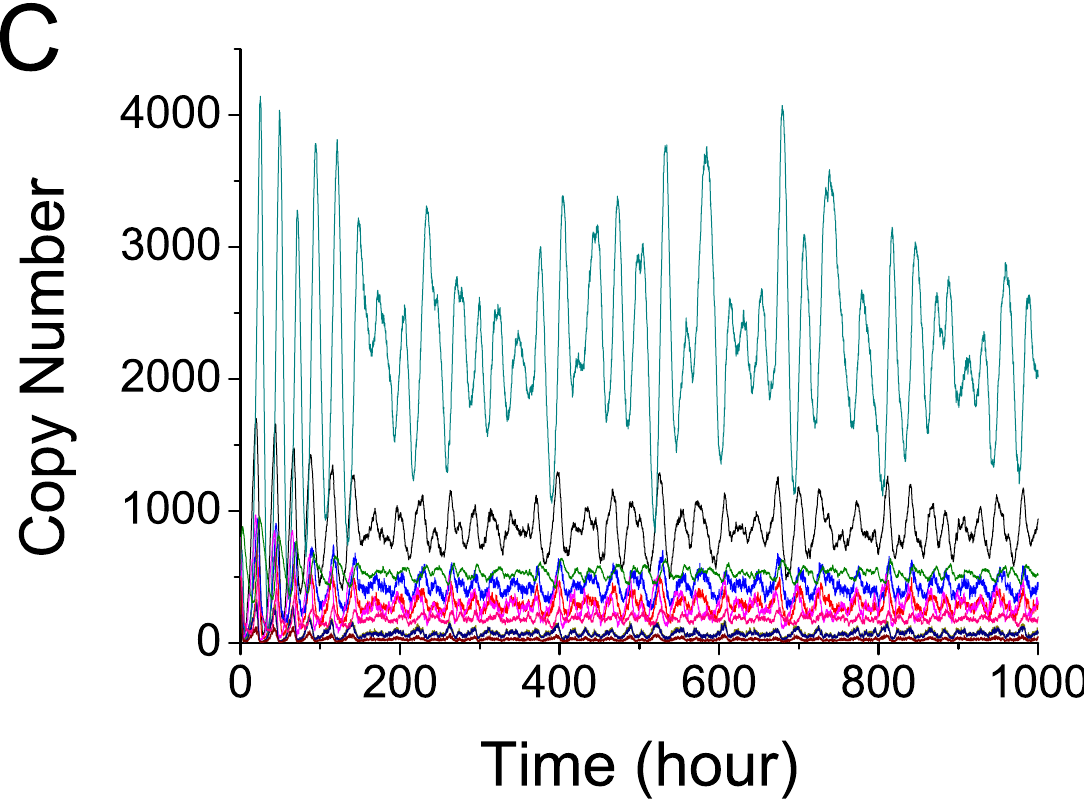}
\end{center}
%\caption{Intrinsic noise is introduced to the original model of
%  Drosophila. Similar phenomena are observed. When cell size is very
%  large, namely about 100 times of a typical cell volume, i.e. $10^5
%  \mu m$ (A), the system would switch from oscillation state to steady
%  state at a random moment due to noise.  As the cell size goes down
%  ($10^4 \mu m$ in B, $10^3 \mu m$ in C), the switching becomes more
%  and more frequent.  A typical cell volumes for Drosophila is about
%  $10^3 \mu m$ (C), which exhibit very noisy dynamics.  }
%\label{noise_Drosophila}
%\end{figure}

\newpage
Figure 8

%\begin{figure*}
\includegraphics[width=\textwidth]{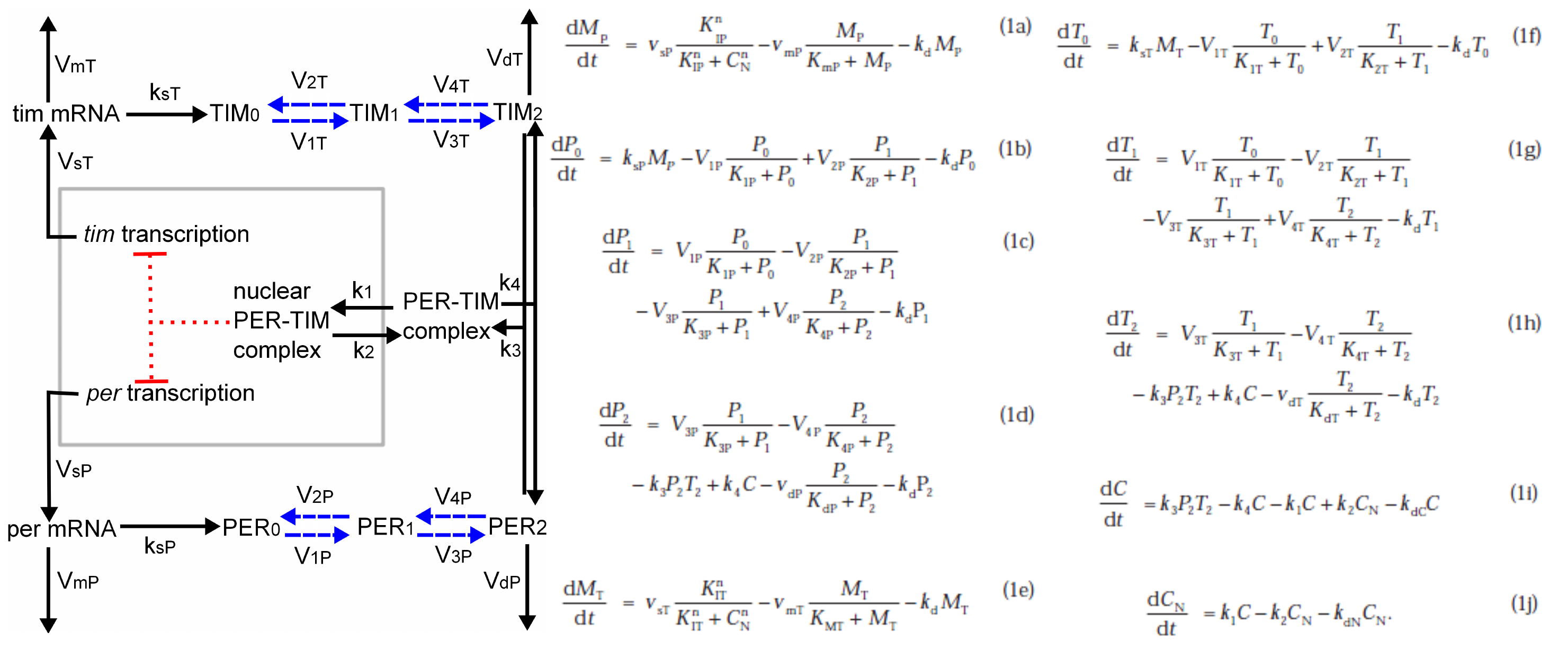}
%\caption{
%The model for Drosophila circadian rhythm
%given in ref \cite{Leloup2,Leloup3}. 
%Negative feedbacks are shown with red, while 
%positive feedback loops are shown with blue.
%The deterministic equation of the model is shown in the 
%right panel, where variables are concentrations of 
%{\sl per} ($M_p$) and tim ($M_T$) mRNAs,
%the PER and TIM with three phosphorylation levels of $P_0$ ($T_0$),
%$P_1$ ($T_1$), and $P_2$ ($T_2$), respectively, 
%the PER-TIM complex $C$, and the nuclear form of the PER-TIM complex
%($C_N$).
%Parameters used are:
%$n = 4$,  $v_{sP}$ = 1.1 nM/h,  $v_{sT}$ = 1 nM/h,  $v_{mP}$ =1.0 nM/h, 
%$v_{mT}$ = 0.7nM/h, $v_{dP}$ = 2.2nM/h, $k_{sP} = k_{sT}$ = 0.9 /h, 
%$k_1$ = 0.8 /h, $k_2$ = 0.2 /h, $k_{3}$ = 1.2/(nM$\cdot$ h), $k_4$ = 0.6 /h, 
%$K_{mP} =K_{mT}$ = 0.2 nM, $K_{IP} =K_{IT} $=1nM, 
%$K_{dP} =K_{dT}$=0.2nM, 
%$K_{1P}=K_{1T}=K_{2P}=K_{2T}=K_{3P}=K_{3T}=K_{4P}=K_{4T}$=2nM, 
%$V_{1P}=V_{1T}$=8nM/h, 
%$V_{2P}=V_{2T}$=1nM/h, $V_{3P}=V_{3T}$=8nM/h, 
%$V_{4P}=V_{4T}$= 1 nM/h, 
%$k_d = k_{dC} = k_{dN} $= 0.01 /h, 
%$V_{dT}$= 1.3nM/h.
%}\label{drosophila}
%\end{figure*}

\end{document}